\documentclass[final,3p,authoryear,12pt]{elsarticle}

\usepackage{graphicx}
\graphicspath{{Plots/}}
\usepackage{amssymb}
\usepackage{url}
\usepackage{multirow}
\usepackage[dvipsnames]{xcolor}
\usepackage{ntheorem}
\usepackage[fleqn]{amsmath}
\usepackage{tablists}
    \restorelistitem 
\usepackage{bm}

\theorembodyfont{\upshape}
\theoremseparator{.}

\usepackage{lineno}
\usepackage{algorithm}
\usepackage{bbm}
\usepackage{natbib}
\usepackage[colorlinks=true,pdfborder={0 0 0}]{hyperref}

\journal{Elsevier}

\begin{document}

\begin{frontmatter}


\title{Assessing the impacts of tradable credit schemes through agent-based simulation}



\author[1]{Renming Liu}
\ead{liure@dtu.dk}
\author[1]{Dimitrios Argyros\corref{corr1}}
\ead{diar@dtu.dk}
\author[1]{Yu Jiang}
\ead{yujiang@dtu.dk}
\author[2]{Moshe E. Ben-Akiva}
\ead{mba@mit.edu}
\author[1]{Ravi Seshadri}
\ead{ravse@dtu.dk}
\author[1]{Carlos Lima Azevedo}
\ead{climaz@dtu.dk}

\address[1]{Department of Technology, Management and Economics, Technical University of Denmark, Denmark}
\address[2]{Department of Civil and Environmental Engineering, Massachusetts Institute of Technology, United States}
\cortext[corr1]{corresponding author}

\begin{abstract}
Tradable credit schemes (TCS) have been attracting interest from the transportation research community as an appealing alternative to congestion pricing, due to the advantages of revenue neutrality and equity. 
Nonetheless, existing research has largely employed network and market equilibrium approaches with simplistic characterizations of transportation demand, supply, credit market operations, and market behavior. Agent- and activity-based simulation affords a natural means to comprehensively assess TCS by more realistically modeling demand, supply, and individual market interactions. We propose an integrated simulation framework for modeling a TCS, and implements it within the state-of-the-art open-source urban simulation platform SimMobility, including: (a) a flexible TCS design that considers multiple trips and explicitly accounts for individual trading behaviors; (b) a simulation framework that captures the complex interactions between a TCS regulator, the traveler, and the TCS market itself, with the flexibility to test future TCS designs and relevant mobility models;
and (c) a set of simulation experiments on a large mesoscopic multimodal network combined with a Bayesian Optimization approach for TCS optimal design.

The experiment results indicate network and market performance to stabilize over the day-to-day process, showing the alignment of our agent-based simulation with the known theoretical properties of TCS. We confirm the efficiency of TCS in reducing congestion under the adopted market behavioral assumptions and open the door for simulating different individual behaviors. We measure how TCS impacts differently the local network, heterogeneous users, the different travel behaviors, and how testing different TCS designs can avoid negative market trading behaviors.

\end{abstract}


\end{frontmatter}


\section{Introduction}
\label{S:1}
Travel demand management aims to reduce road traffic congestion by altering patterns of trip-making, activity participation and mode, departure time and, route choices, with congestion pricing being one of the most prominent measures. 
Although traditionally, congestion pricing has been popular in both theory and practice, with empirically demonstrable gains in social welfare \citep{lindsey2006economists}, it often receives political and social resistance as it is perceived as a tax \citep{de2020tradable}. A tradable credit scheme (TCS) is a form of quantity control, which offers an appealing alternative that has the potential of addressing the issues of political and social opposition if properly designed. 

In a typical TCS system, a regulator provides an initial endowment of mobility credits to potential travelers. The credits can be bought and sold in a market monitored by the regulator at a price determined by demand and supply \citep{grant2014role}. Tradable credits provide two key advantages over congestion pricing without refunding: 1) there is no net financial flow from road users to the regulator, making it revenue-neutral, and potentially, less susceptible to political opposition; 2) in theory, it can achieve any desired equity distribution through the initial credit allocation, and it enables the regulator to directly control quantity (e.g., of road use) \citep{fan2013tradable}. Since the early work of \cite{goddard1997using,verhoef1997tradeable}, there has been a surge in academic attention towards TCS. For a comprehensive review of different TCS conceptualizations, designs, implementations, and credit distributions the reader is referred to \cite{fan2013tradable, provoost2023design, servatius2023trading}. \cite{grant2014role} review focuses on the studies that formulate mathematical programming models to investigate the user and market equilibrium considering transaction costs, fixed/elastic demand, and homogeneous/heterogeneous travelers. \cite{dogterom2017tradable} summarizes the empirical findings in individual behaviors under different TCS systems, including loss aversion, mental accounting and decision-making under uncertainty. Subsequent extensions include consideration of multi-period TCS \citep{miralinaghi2016multi,miralinaghi2018multi}, cyclic TCS \citep{xiao2019promoting}, comparative analysis with congestion pricing \citep{de2018congestion,seshadri2022congestion}, dynamics of the credit price \citep{ye2013continuous,guo2019tradable,balzer2022modal,liu2022managing}, detailed individual market behaviors (namely, selling and buying) \citep{tian2019understanding, chen2023market, hamm2023mobilitycoins}, transaction costs in TCS \citep{zhang2021tradable,fan2022managing}, multimodality in TCS \citep{balzer2023dynamic},  and TCS for parking management \citep{liu2014novel,xiao2021tradable,bao2022tradable, alogdianakis2024development}.

Despite the large body of research on TCS, there are several challenges that need to be addressed before real-world deployments are possible.
First, most previous studies use simplistic demand and supply models for assessment and employ static equilibrium approaches to model the credit market (with the exception of \cite{ye2013continuous,guo2019tradable, balzer2022modal,liu2022managing} where the flow and credit price dynamics are captured). Thus, few attempts are made to model the disaggregate behavior of individuals within the market, let alone the modeling of behavioral responses such as loss aversion, budgeting, learning effects, etc. \citep{dogterom2017tradable}. Second, there is little research on the design of the credit system itself, namely, aspects such as credit allocation, trading, market regulation and transaction costs \citep{nie2012transaction, fan2022managing, lessan2022credit}, and the impacts on the system efficiency and individual behavior in the market. 

To summarize, our current understanding of the impacts and operations of a TCS stands to benefit from a framework that can capture the complex interactions of the different dimensions of user behavior, the regulator actions and the market design, so as to provide comprehensive assessments of different TCS. In this paper, we contribute to the existing literature 
by: (a) proposing a flexible simulation-based framework to model a TCS including the explicit consideration of the disaggregate behavior of users and their interactions within the market and with the regulator (the framework also proposes a model of individual selling behavior that extends the one proposed in \cite{chen2023market} to consider multiple trips), (b) implementing the framework in a modular and extensible manner in the state-of-the-art urban simulator SimMobility \citep{adnan2016simmobility}, allowing for the detailed simulation of the operations of a TCS system, and (c) conducting simulation experiments using a prototypical city to showcase the functionality of the proposed framework and to yield insights into the impacts of TCS in mitigating congestion, TCS design and market behavior. 

SimMobility is an integrated agent- and activity-based simulation platform using a multi-scale framework comprised of three primary modules at different temporal scales: Long-term (year-to-year) \citep{adnan2016simmobility}, mid-term (day-to-day) \citep{lu2015simmobility}, and short-term (within day) \citep{azevedo2017simmobility}. Specifically, SimMobility mid-term simulates agents' behavior, including their activity and travel patterns and the movement of vehicles and travelers. This study extends the mid-term module with new functionalities that can simulate the operations of a complex TCS system.


The rest of this paper is structured as follows. In Section \ref{S:2}, we introduce the TCS design and the proposed market behavior model, which includes deriving an optimal selling strategy considering multiple daily trips. The architecture of the SimMobility mid-term module and its core design components are discussed in Section \ref{S:simdesign}. Section \ref{sec:Opt} describes the optimization formulation to determine the credit charging scheme (toll profile in credits). Next, simulation results are discussed in Section \ref{S:Results}, followed by conclusions and directions for future research in Section \ref{5_S:Con}.

\section{TCS framework}
\label{S:2}
The TCS we consider in our simulation framework extends the design presented in \cite{brands2020tradable} and \cite{chen2023market} to incorporate multiple trips within a day. The framework is generic and allows for various toll designs including distance-based, area-based and cordon-based schemes. For the sake of completeness, we provide a summary of the TCS/market design in Section \ref{sec:TCSDesign} and describe our proposed extensions for agent-based and network simulation in Section \ref{sec:MB}. 
\subsection{TCS design}\label{sec:TCSDesign}
Here, the TCS has the following features:

\textit{Credit allocation}: We use a `continuous' allocation approach wherein the regulator gives out credits to every traveler at a certain rate over the entire day. Compared to a lump-sum allocation \citep{brands2020tradable}, the `continuous' allocation can potentially prevent concentrated trading activities due to credit expiration and provides additional degrees of freedom to the regulator to intervene in the market \citep{chen2023market}. The lump-sum allocation often used in the literature \citep{brands2020tradable} is a special case of the `continuous' allocation where the allocation interval is relatively large (for example, one day).

\textit{Credit expiration}: Each credit has a certain initial lifetime specified by the regulator. 
The expiration of credits can avoid speculative behavior in the market, such as credit stocking and banking \citep{de2020tradable}.

\textit{Credit transactions}: All travelers are assumed to trade directly with the regulator, who guarantees all buying and selling requests. Such a `traveler-to-regulator' trading mechanism is beneficial in reducing transaction costs associated with information acquisition, negotiation, and other peer-to-peer trading and auctioning features \citep{brands2020tradable}. Thus, travelers who are short of credits can only buy the remaining credits needed at departure and for immediate use. Furthermore, when a traveler wishes to sell credits to the regulator, it has to sell all credits in its credit account. Finally, travelers are not allowed to sell and buy credits at the same time.

\textit{Credit toll}: When the toll is active, travelers have to pay the toll in credits to use the road network. The toll in credits charged by the regulator is dynamic and varies by time of day but is invariant across days. It can further be either distance-based, area-based, cordon-based or a combination of these. Note that an upfront credit toll at departure is used for all trips.

Each traveler possesses a credit account, whose balance evolves over time. Let $r$ denote the allocation rate of credits, $l$ be the credit lifetime, and $x_n^d(t)$ be traveler $n$'s credit account balance at time $t$ on day $d$. The maximum number of credits accumulated in the account is $l\cdot r$. Once an account has reached this state (referred to as `\textit{full wallet}' hereafter), the balance does not change in the absence of selling or traveling, as every time a new credit is acquired, the oldest credit expires. In contrast, when the account balance is smaller than $l\cdot r$, it increases by $r\cdot \Delta t$ in a time interval $\Delta t$. Let $g(t)$ denote the credit toll at time $t$ and $t_{n,i,d}^{dep}$ be the departure time of traveler $n$'s $i$th trip on the day $d$. Note that at time $t$ on the day $d$, traveler $n$ can perform one and only one of the three actions that affect the account balance: start a trip (and possibly buy credits), do nothing, or sell all credits. We can then derive the account balance at time $t+\Delta t$ as follows:
\begin{enumerate}
  \item[A.] Start a trip when $t=t_{n,i,d}^{dep}$
  \begin{itemize}
    \item if $x_n^d(t)\geq g(t)$, traveler $n$'s account balance at time $t+\Delta t$ is given by:
    \begin{equation}\label{account1}
    x_n^d(t+\Delta t) = \min (x_n^d(t)-g(t)+r\cdot \Delta t,~l\cdot r).
    \end{equation}
    \item if $x_n^d(t)\leq g(t)$, traveler $n$ needs to buy $g(t)-x_n^d(t)$ credits, and the account balance at time $t+\Delta t$ is:
    \begin{equation}\label{account2}
    x_n^d(t+\Delta t) = r\cdot \Delta t,
    \end{equation}
    as all credits in the account and bought credits are consumed for the trip.
  \end{itemize}
  \item[B.] Do nothing. The account balance $x_n^d(t+\Delta t)$ becomes:
  \begin{equation}\label{account3}
    x_n^d(t+\Delta t) = \min (x_n^d(t)+r\cdot \Delta t,~l\cdot r).
    \end{equation}
  \item[C.] Sell all credits $x_n^d(t)$, then:
  \begin{equation}\label{account4}
    x_n^d(t+\Delta t) = r\cdot \Delta t.
    \end{equation}
\end{enumerate}

\textit{Credit price}: As credits are bought and sold in a market, the price is determined endogenously by credit demand-supply interactions \citep{yang2011managing}. We use the price adjustment mechanism from \cite{liu2022managing} where the credit price is adjusted from day to day but fixed within a day. Specifically, the price on day $d$, $p_d$, increases or decreases proportionally to the previous day’s excess credit consumption $Z_d$, defined as the difference between the total numbers of bought and sold credits by all travelers. Thus,
\begin{equation}\label{price}
p_{d+1}=\max\{p_{d}+ k* Z_d,~0\},
\end{equation}
where $k$ is the price adjustment parameter. When the credit demand exceeds supply, the market price increases, and vice versa.

\subsection{User's market behavior}\label{sec:MB}
We assume that the regulator levies a transaction fee on each transaction (buying and selling), which is composed of two parts, a fixed fee and a fee proportional to the value of traded credits. Let $f_b,~f_s \geq 0$ and $\hat{f}_b,~\hat{f}_s \geq 0$ denote the fixed and proportional fees for selling and buying transactions, respectively. We can write the revenue $S$ of selling $y$ credits with transaction fees on day $d$ as:
\begin{equation}\label{sell_rev}
S(y|p_d, f_s, \hat{f}_s )=y\cdot p_d\cdot(1-\hat{f}_s)-f_s.
\end{equation}
Similarly, the cost $B$ of buying $y$ credits with transaction fees on day $d$ is given by:
\begin{equation}\label{buy_cost}
B(y|p_d, f_s, \hat{f}_s )=y\cdot p_d\cdot(1+\hat{f}_b)+f_b.
\end{equation}

As described in Section \ref{sec:TCSDesign}, a buying transaction may only happen when a traveler starts its trip, while a selling transaction can occur at every time interval. Here we adopt the heuristic selling model from \cite{chen2023market}, where an individual's selling strategy is developed by considering only one trip per day (i.e., the morning commute trip). We extend this model to accommodate multiple trips and activities in a single day. 
At time $t$ on day $d$, assume traveler $n$ has $I$ upcoming trips at times denoted by $t_{n,i,d}^{dep},~i=1,2,\dots,I$; the expected profit of selling all credits at time $t<t_{n,1,d}^{dep}$ on day $d$ without further selling, $P_n^d(t)$, can be written as follows,
\begin{equation}
\begin{aligned}
\label{profit}
    P_n^d(t) =& S(x_n^d(t)) - \sum_{i=1}^I \mathbbm{1}(g(t_{n,i,d}^{dep})>x_n^d(t_{n,i,d}^{dep}))\cdot B(g(t_{n,i,d}^{dep})-x_n^d(t_{n,i,d}^{dep}))\\
    =& x_n^d(t)\cdot p_d \cdot (1-\hat{f}_s)-f_s\\
    &- \sum_{i=1}^I\mathbbm{1}(g(t_{n,i,d}^{dep})>x_n^d(t_{n,i,d}^{dep}))\cdot \Big(\big(g(t_{n,i,d}^{dep})-x_n^d(t_{n,i,d}^{dep})\big)\cdot p_d\cdot(1+\hat{f}_b)+f_b \Big),
\end{aligned}
\end{equation}
where the expected account balance at the departure time of trip $i$ is given by (recall that traveler $n$ is assumed to have sold all of her credits at time $t$),
\begin{equation}\label{acc_bal}
x_n^d(t_{n,i,d}^{dep})=
\left\{
\begin{aligned}
     & \min\{(t_{n,1,d}^{dep}-t)\cdot r,~l\cdot r\},\quad i=1,\\
     &\min\{\max\{x_n^d(t_{n,i-1,d}^{dep})-g(t_{n,i-1,d}^{dep}),0\}+(t_{n,i,d}^{dep}-t_{n,i-1,d}^{dep})r,l\cdot r\}, i=2,\dots,I,\\
\end{aligned}
\right.
\end{equation}
where $t<t_{n,1,d}^{dep}$. Note that the buying cost is only incurred when the toll for trip $i$ on day $d$, $g(t_{n,i,d}^{dep})$, is larger than the account balance at the time of departure of trip $i$, $x_n^d(t_{n,i,d}^{dep})$, as written by the indicator function in equation \eqref{profit}. A critical assumption made for computing the selling profit is that there is no further selling until the last trip of the trip chain, which consists of $I$ trips on day $d$ and duplicated $I$ trips on day $d+1$. Given the `\textit{sell all credits}' rule and the presence of transaction fees, it is a justifiable assumption that allows us to derive the optimal selling strategy analytically. 

A traveler checks for two conditions when it is making a selling decision: i) the selling profit $P_n^d(t)$ must be positive, and ii) the profit from selling now must be higher than that obtained from selling at any other time until the next departure (i.e., $t'<t_{n,1,d}^{dep}$), which can be determined by examining the sign of the derivative of $P_n^d(t)$ with regard to $t$. Thus, we can compute both the profit value and its derivative to obtain the optimal selling strategy. We analyze the derivative of profit function \eqref{profit} in cases with different outcomes of the indicator functions. For ease of demonstration, here we 
simplify $\mathbbm{1}(g(t_{n,i,d}^{dep})>x_n^d(t_{n,i,d}^{dep}))$ as $\mathbbm{1}(t_{n,i,d}^{dep})$, and take the case of two trips, $1$ and $2$,  as an example:
\begin{enumerate}
    \item[A.] If $g(t_{n,1,d}^{dep})\leq x_n^d(t_{n,1,d}^{dep})$ and $g(t_{n,2,d}^{dep})\leq x_n^d(t_{n,2,d}^{dep})$, i.e., $\mathbbm{1}(t_{n,1,d}^{dep})=\mathbbm{1}(t_{n,2,d}^{dep})=0$, we have
    \begin{equation}\label{prof1}
    P_n^d(t) = x_n^d(t)\cdot p_d \cdot (1-\hat{f}_s)-f_s,
    \end{equation}
    and the derivative,
    \begin{equation}\label{derive1}
    \frac{\partial P_n^d(t)}{\partial t}=
    \left\{
    \begin{aligned}
         & 0,\quad \text{if}~x_n^d(t)=l\cdot r,\\
         &r\cdot p_d \cdot (1-\hat{f}_s)>0,\quad \text{otherwise}.\\
    \end{aligned}
    \right.
    \end{equation}
    This implies that the profit will increase until the account balance reaches the maximum amount of $l\cdot r$ (i.e., full wallet). Though the derivative is 0 at the full wallet state, there is no incentive for a traveler to defer the selling since the oldest credits start to expire. Hence, the selling (of all credits) happens at the full wallet state. It is noteworthy that when the fixed transaction fee $f_s=0$, the selling profit at full wallet state is the same as the profit of any selling strategy that avoids credit expiration (e.g., selling every time new credits are allocated). A positive transaction fee can prevent frequent selling with a small number of credits \citep{brands2020tradable,chen2023market}.

    \item[B.] If $g(t_{n,1,d}^{dep})>x_n^d(t_{n,1,d}^{dep})$ and $g(t_{n,2,d}^{dep})>x_n^d(t_{n,2,d}^{dep})$, i.e., $\mathbbm{1}(t_{n,1,d}^{dep})=\mathbbm{1}(t_{n,2,d}^{dep})=1$, we have
    \begin{equation}\label{prof2}
    \begin{aligned}
        P_n^d(t) =& x_n^d(t)\cdot p_d \cdot (1-\hat{f}_s)-f_s\\
        &-\Big(\big(g(t_{n,1,d}^{dep})-x_n^d(t_{n,1,d}^{dep})\big)\cdot p_d\cdot(1+\hat{f}_b)+f_b\Big)\\
        &-\Big(\big(g(t_{n,2,d}^{dep})-x_n^d(t_{n,2,d}^{dep})\big)\cdot p_d\cdot(1+\hat{f}_b)+f_b\Big),
    \end{aligned}
    \end{equation}
    and the derivative as
    \begin{equation}\label{derive2}
    \frac{\partial P_n^d(t)}{\partial t}=
    \left\{
    \begin{aligned}
         & -2r\cdot p_d\cdot(1+\hat{f}_b)<0,\quad \text{if}~x_n^d(t)=l\cdot r,\\
         &r\cdot p_d \cdot (1-\hat{f}_s)-2r\cdot p_d\cdot(1+\hat{f}_b)<0,\quad \text{otherwise},\\
    \end{aligned}
    \right.
    \end{equation}
    which is always negative. This implies that the profit will decrease and thus selling at the current time is the optimal strategy if the profit is positive.
    \item [C.] If $g(t_{n,1,d}^{dep})>x_n^d(t_{n,1,d}^{dep})$ and $g(t_{n,2,d}^{dep})\leq x_n^d(t_{n,2,d}^{dep})$, i.e., $\mathbbm{1}(t_{n,1,d}^{dep})=1$ and $\mathbbm{1}(t_{n,2,d}^{dep})=0$, we have
    \begin{equation}\label{prof3}
    \begin{aligned}
        P_n^d(t) =& x_n^d(t)\cdot p_d \cdot (1-\hat{f}_s)-f_s-\Big(\big(g(t_{n,1,d}^{dep})-x_n^d(t_{n,1,d}^{dep})\big)\cdot p_d\cdot(1+\hat{f}_b)+f_b\Big)
    \end{aligned}
    \end{equation}
    and the derivative as
    \begin{equation}\label{derive3}
    \frac{\partial P_n^d(t)}{\partial t}=
    \left\{
    \begin{aligned}
         & -r\cdot p_d\cdot(1+\hat{f}_b)<0,\quad \text{if}~x_n^d(t)=l\cdot r,\\
         &r\cdot p_d \cdot (1-\hat{f}_s)-r\cdot p_d\cdot(1+\hat{f}_b)\leq 0,\quad \text{otherwise}.\\
    \end{aligned}
    \right.
    \end{equation}
    \item[D.] If $g(t_{n,1,d}^{dep})\leq x_n^d(t_{n,1,d}^{dep})$ and $g(t_{n,2,d}^{dep})> x_n^d(t_{n,2,d}^{dep})$, i.e., $\mathbbm{1}(t_{n,1,d}^{dep})=0$ and $\mathbbm{1}(t_{n,2,d}^{dep})=1$, we have
    \begin{equation}\label{prof4}
    \begin{aligned}
        P_n^d(t) =& x_n^d(t)\cdot p_d \cdot (1-\hat{f}_s)-f_s-\Big(\big(g(t_{n,2,d}^{dep})-x_n^d(t_{n,2,d}^{dep})\big)\cdot p_d\cdot(1+\hat{f}_b)+f_b\Big).
    \end{aligned}
    \end{equation}
    and the derivative as
    \begin{equation}\label{derive4}
    \frac{\partial P_n^d(t)}{\partial t}=
    \left\{
    \begin{aligned}
         & -r\cdot p_d\cdot(1+\hat{f}_b)<0,\quad \text{if}~x_n^d(t)=l\cdot r,\\
         &r\cdot p_d \cdot (1-\hat{f}_s)-r\cdot p_d\cdot(1+\hat{f}_b)\leq 0,\quad \text{otherwise}.\\
    \end{aligned}
    \right.
    \end{equation}
    For both cases $C$ and $D$, the derivative is negative when the proportional transaction fees $\hat{f}_s$ and $\hat{f}_b$ are positive, i.e., a traveler should thus sell all credits now. Without transaction fees, the derivative becomes 0 when the account balance is not full, and the selling profits now or later are the same. Positive fixed transaction fees make selling at full wallet beneficial for the sake of transaction cost minimization, and positive proportional transaction fees make immediate selling preferable as the derivative becomes negative.
\end{enumerate}

We can easily extend the above analysis for cases with multiple trips. For all future $I$ trips:
\begin{enumerate}
    \item[A.] if all the tolls in credits are smaller than the corresponding expected account balances, the derivative of profit function is
    \begin{itemize}
        \item 0, if the current account balance is full,
        \item positive, otherwise.
    \end{itemize}
    \item[B.] if more than one toll in credits is larger than the corresponding expected account balance, the derivative of profit function is always negative.
    \item[C.] if there is only one toll in credit larger than the corresponding expected account balance, the derivative of profit function is
    \begin{itemize}
        \item 0, if the proportional transaction fees are 0,
        \item negative, otherwise.
    \end{itemize}
\end{enumerate}

Therefore, considering positive transaction fees, we can summarize the selling strategy for traveler $n$ at time $t$ on the day $d$ as follows:
\begin{itemize}
    \item the profit must be positive, $P_n^d(t)>0$.
    \item if the account balance is full, i.e., $x_n^d(t)=l\cdot r$, then sell now.
    \item if $g(t_{n,i,d}^{dep})\geq x_n^d(t_{n,i,d}^{dep})~\exists i=1,2,\dots,I$, then sell now.
    \item if $g(t_{n,i,d}^{dep})<x_n^d(t_{n,i,d}^{dep})~\forall i=1,2,\dots,I$, then do not sell.
\end{itemize}

Note that this selling strategy is essentially myopic since no further selling is considered. That can be considered using a more complex dynamic programming formulation, however, we defer that into future research.

\subsection{Interactions of travelers, the market and the regulator}\label{sec:interaction}
An operational system for TCS contains three types of actors or agents: travelers, the regulator, and a market platform. The regulator gives out credits to all travelers at a certain rate $r$. At a given time $t$, travelers determine their roles as either sellers or buyers depending on the account balance and the credit needs associated with their mobility choices.

We design a market platform agent, which handles, collects and records all transactions, and interacts with both travelers and the regulator. Travelers need to send trading requests through the market platform to the regulator and proceed with the transactions after approval. That entails several checks, for instance, a buying request might be rejected if the cap of credit usage has been reached, and a selling request could be rejected if the regulator's revenue is lower than a predetermined threshold. That process also underlies the adjustment of credit price, which again is not the responsibility of the regulator, but a result of credit demand-supply interactions, using the collected transaction information on a daily basis.

\begin{figure}[H]
    \centering
    \includegraphics[width=0.95\textwidth]{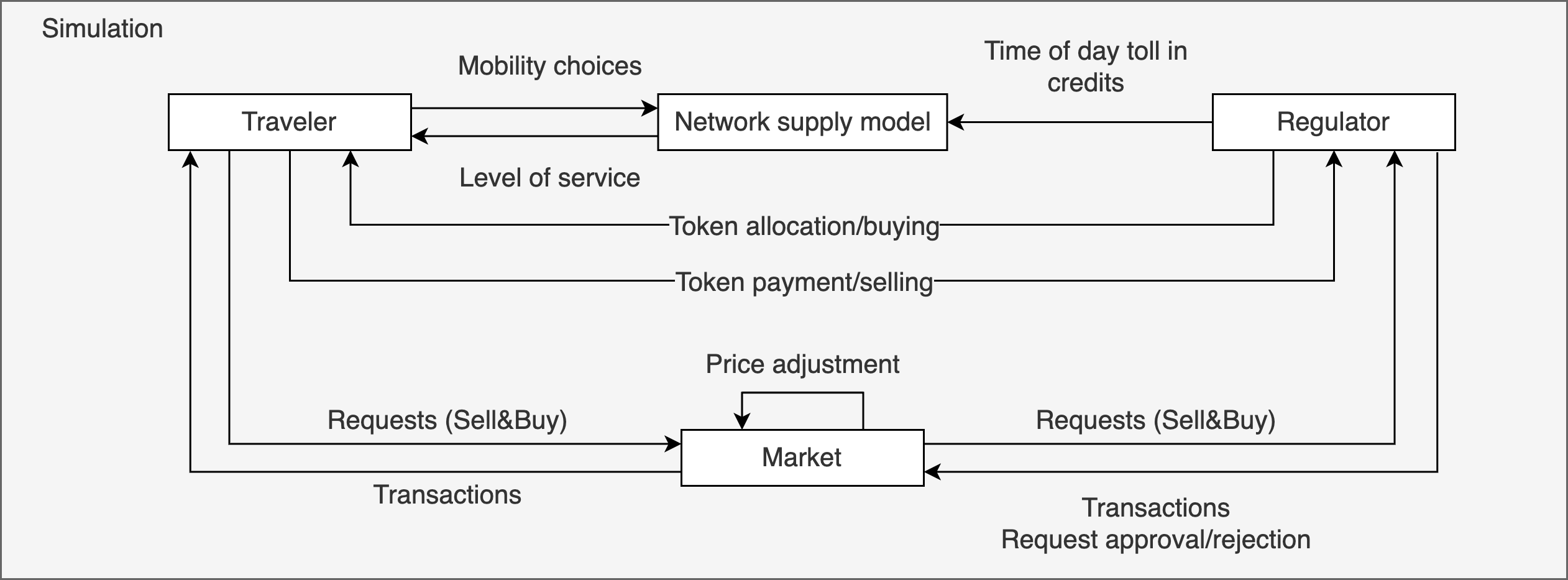}
    \caption{Interactions between the travelers, regulator, and market}
    \label{interaction}
\end{figure}

The overall framework of interactions is shown in Figure \ref{interaction}. In Section \ref{sec:class_design}, we will introduce in detail the design of classes, functions, and the information synchronization established based on the interaction framework in the simulation platform.

\section{Simulation platform and design}
SimMobility is an agent- and activity-based multi-level urban transportation simulation platform with an open-source codebase. SimMobility consists of three simulators at different temporal dimensions: The Long-term module simulates the evolution of population synthesis, residential and workplace locations, vehicle ownership, density, and land use distribution over a long period (typically from months to years) \citep{zhu2018integrated}; the Mid-term module covers daily activity scheduling, mode, route, destination and departure time choices on a multi-modal network \citep{lu2015simmobility}; finally, the Short-term module captures the vehicle behaviors on the road network at a high spatial-temporal resolution (in the order of milliseconds), including lane-changing, braking and accelerating, and device-to-device communications, and synthesizes traffic control and management systems \citep{azevedo2017simmobility}. These three simulators are integrated in a way that decisions made in longer terms (e.g., Long-term) are inputs to decision-making in shorter terms (e.g, Mid-term), which, in turn, affect longer terms by providing supporting information (e.g., accessibility measures) \citep{adnan2016simmobility}. Thus, the design and development of a TCS simulation framework within SimMobility provides a wide range of assessment capabilities from a demand-side perspective compared to other existing frameworks \citep{anda2017transport}. 
\label{S:simdesign}
\subsection{Overview of SimMobility Mid-term}
\begin{figure}[htbp]
    \centering
    \includegraphics[width=0.7\textwidth]{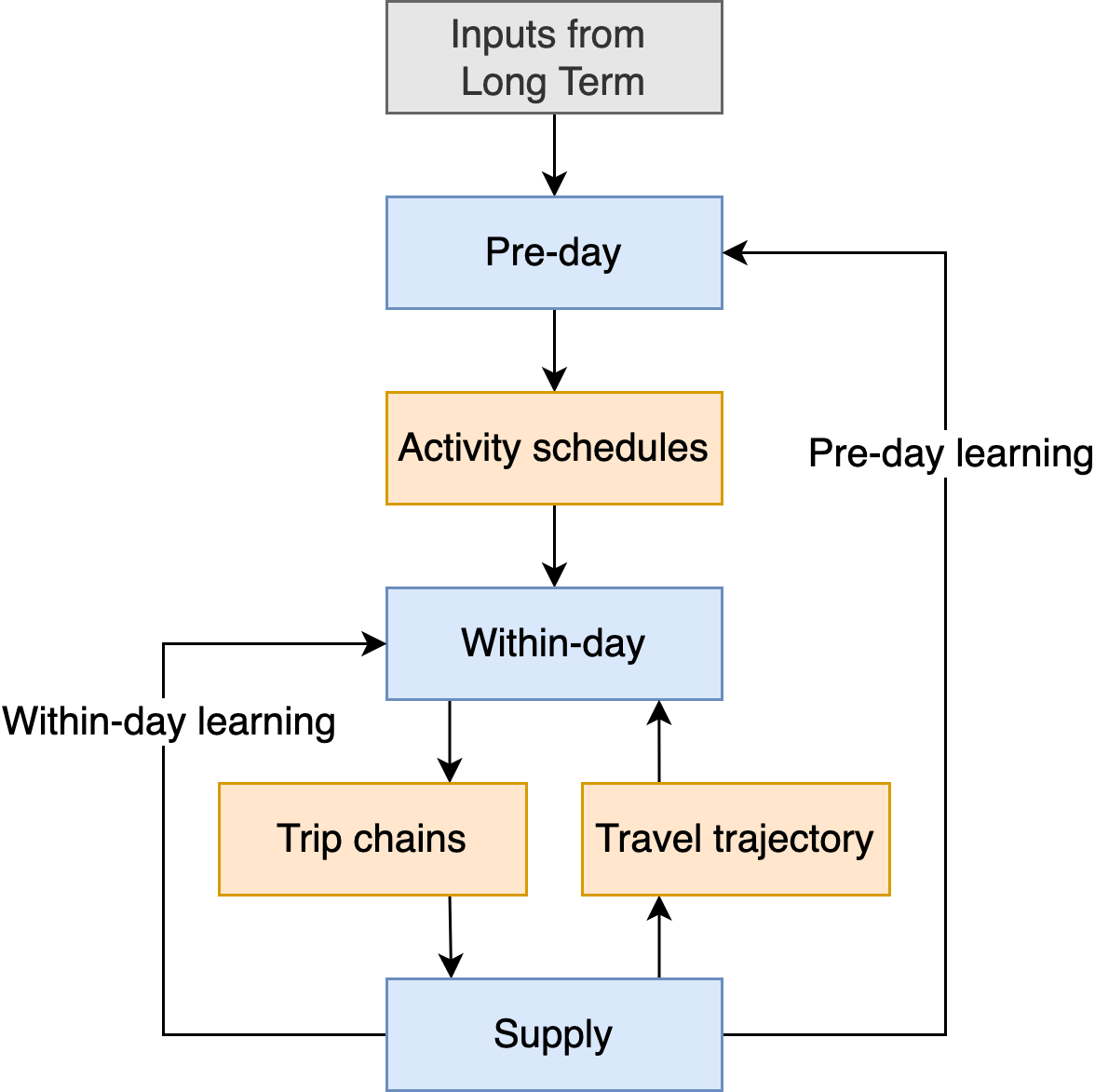}
    \caption{SimMobility mid-term framework}
    \label{midterm}
\end{figure}
In this paper, we focus our development on the extension of SimMobility Mid-term for a wide range of TCS simulations. SimMobility Mid-term\footnote{https://github.com/smart-fm/simmobility-prod\label{SMgithub}} is an open-source mesoscopic simulator that models agents' daily activities at the individual level on the demand side and simulates the movement at the mesoscopic level on the supply side \citep{adnan2016simmobility}. SimMobility Mid-term comprises three main modules: \textit{Pre-day},\textit{Within-day}, and \textit{Supply}, shown as Figure \ref{midterm}.

The \textit{Pre-day} module adopts the Day Activity Schedule approach \citep{bowman2001activity} to predict the individual daily activity schedule, including activity type and number, activity duration and time of day (by half-hour time intervals), activity locations, and modes (See \cite{siyu2015activity} for more details on the \textit{Pre-day} model). Given the information from the \textit{Pre-day} output, the \textit{Within-day} module extracts the trip chains and simulates travelers' departure time and route choice decisions before their trips as well as en-route decisions. The \textit{Supply} module, which is a mesoscopic multi-modal traffic simulator, takes in the trip chains and simulates the actual movement trajectories. In the experiments in Section \ref{S:Results}, we assume that the \textit{Pre-day} activity schedules are fixed and focus on the \textit{Within-day} module, which is described in Section \ref{sec:dem_ext}.

\subsection{Main components design}\label{sec:class_design}
SimMobility is highly modularized and extensible so that we can expand the existing class definitions (of agents and entities, see \cite{adnan2016simmobility}) to add functionality via new classes, class attributes, operations, and associations between classes. 

The base class used in SimMobility is termed \textit{`Entity'}, which encapsulates fundamental attributes such as its id, and functions such as the \textit{update} function, which is called at every time step of the simulation. An entity may also be a message handler that can receive and react to messages coordinated by a messaging system. Derived from \textit{`Entity'}, class \textit{`Agent'} forms the base for all SimMobility Mid-term agents, including persons, traffic lights, loop-detectors, bus-stop agents, etc. The reader is referred to both \cite{adnan2016simmobility} and footnote \ref{SMgithub} for further details on SimMobility architecture. To establish the operational system for TCS, five main components need to be added to the SimMobility Mid-term simulator: individual TCS accounts, a regulator, a market, the credits and all its related transactions.

\begin{figure}[htbp]
    \centering
    \includegraphics[width=0.85\textwidth]{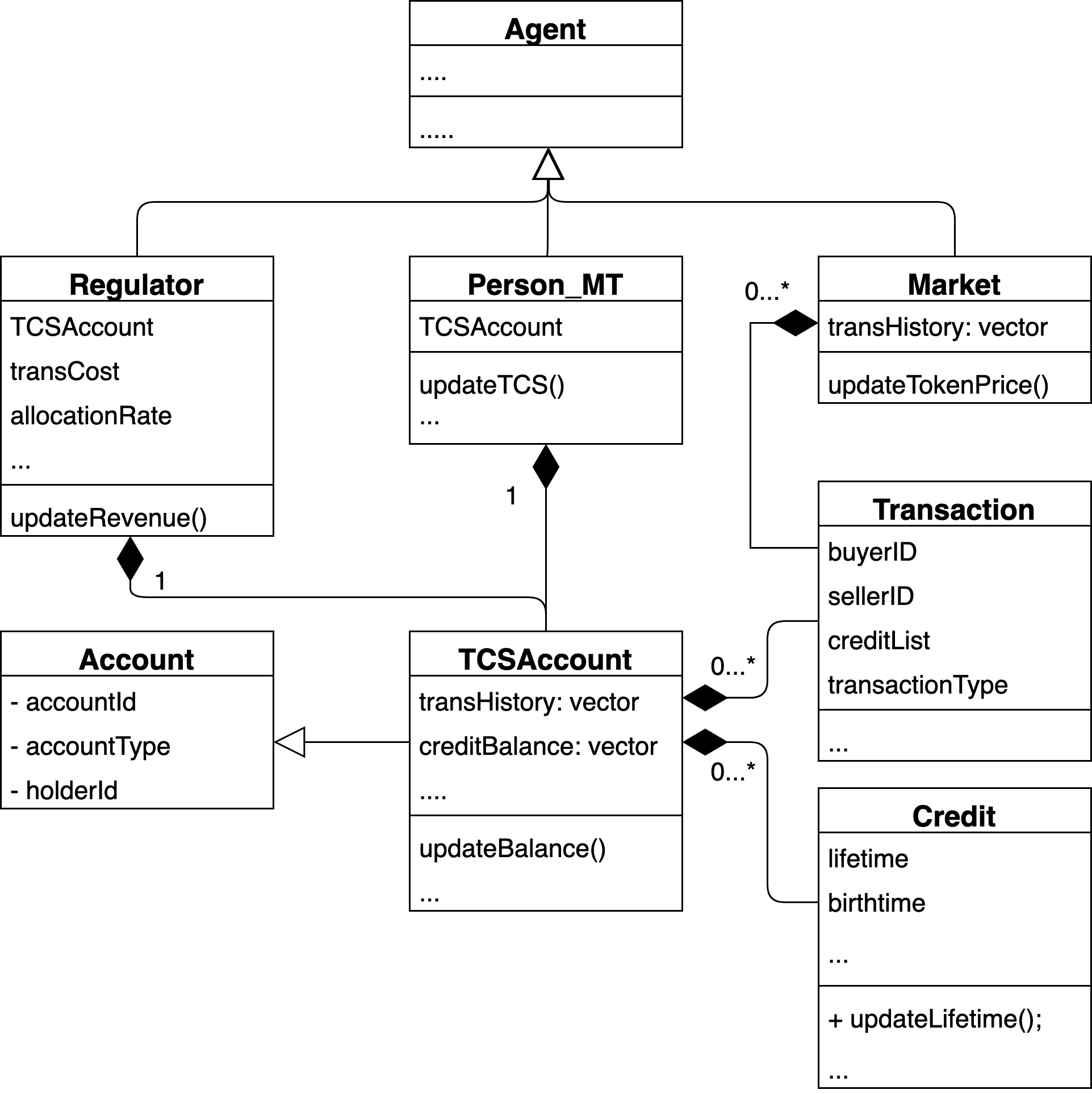}
    \caption{Class diagram of the TCS framework in SimMobility}
    \label{class}
\end{figure}

Figure \ref{class} illustrates the proposed TCS system structure by a class diagram containing each class's main attributes and operations. We define a generic '\textit{Account}' class with basic attributes, including the account ID, account type, and owner ID. Different account types can be used for the purposes of simulating different policies (e.g., user account used for subscriptions and product choices in a congestion pricing or Mobility-as-a-Service setting \citep{kamargianni2019incorporating}). In the case of TCS, the account inherits from the base account class and contains the active credits and transaction records as vectors, as well as auxiliary variables such as numbers of used/bought/sold credits, transaction type, and numbers, and current predicted selling profit. A transaction object is generated when credits are traded or allocated, recording the transaction type (buy, sell, use, and allocation), buyer and seller IDs, and the list of credits. Credits are generated by the regulator and have a birth time and lifetime, which is updated as the simulation proceeds. The new regulator and market, as well as the existing traveler (Person MT) inherit from the base agent class in SimMobility. The regulator has its own TCS account, specifies the allocation rate and transaction costs, and records the revenue from trading with travelers. Each traveler also has her own account and updates it every time a transaction or expiration of credits occurs. The market collects information on all transactions between travelers and the regulator and adjusts the credit price accordingly.

\begin{figure}[htbp]
    \centering
    \includegraphics[width=0.85\textwidth]{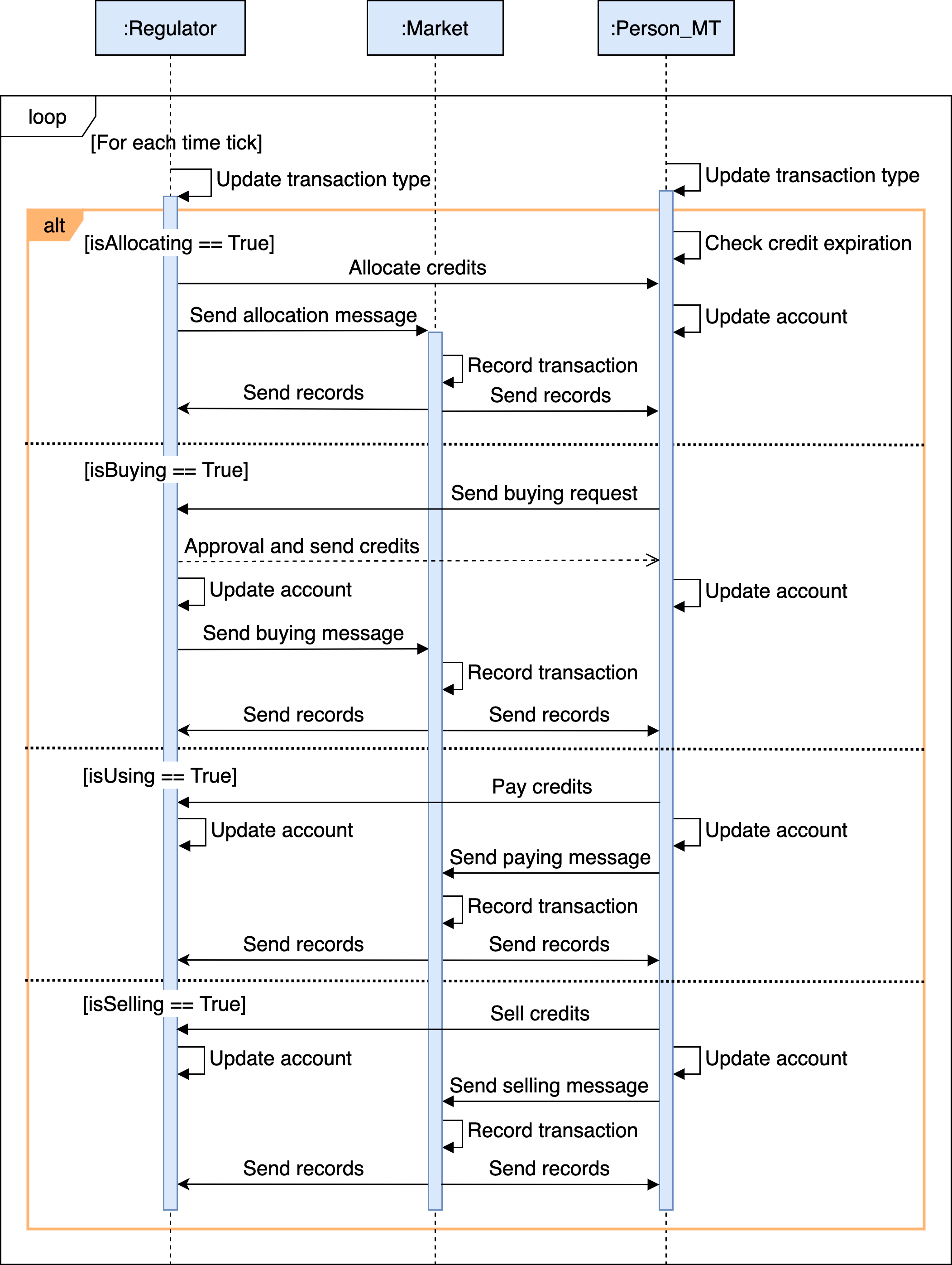}
    \caption{Sequence diagram for information exchanges under the TCS framework in SimMobility}
    \label{sequence}
\end{figure}

Following the interactions described in Section \ref{sec:interaction}, the design of communications among the regulator, market and travelers is shown in the sequence diagram Figure \ref{sequence}. At each time step, the regulator and traveler check for the occurrence of the four possible types of transactions in sequence, specifically, credit allocation first, followed by buying (if needed) and using transactions, and finally selling of credits. The 'continuous' allocation is discretized for a specified time interval, as small as the simulation time step itself (to be optimized for computational efficiency). An allocation is triggered when the current time equals the start time of such predetermined allocation time interval. The regulator sends the credits to all travelers and broadcasts the information to the market, which generates a transaction record and sends it to both the regulator and travelers. Travelers will update their account balances after receiving the credits. Note that in the same time tick, the credit expiration check is automatically conducted (it is clear that a credit can only expire in the time interval of allocation as long as the credit lifetime is a multiple of the allocation interval, and such a design has generality to some extent and can avoid unnecessary computations). For the buying and using case, the use of credits is triggered by the departure time of a trip, and the buying of credits is triggered only when the traveler is short of credits. For a buying transaction, a traveler sends the buying request stating the number of credits, and the regulator sends back the credits upon approval of the request, sending the trading information to the market at the same time. Both the regulator and traveler update their account and receive the transaction record from the market after the trade. Similarly, in the case of using and selling credits, travelers transfer credits to the regulator, and the market records the transaction information. After a certain period, the market will use the aggregated information to adjust the credit price based on the rule specified in Section \ref{sec:TCSDesign}. In this paper, we set the period as a day and defer the within-day price adjustment to future tests.

All the communication of information happens through the SimMobility messaging system \citep{adnan2016simmobility}, which collects the messages from all senders and distributes them to related receivers. The above architecture is generic enough and can be easily extended to handle any possible TCS designs, particularly peer-to-peer (P2P) trading mechanisms. A P2P trading mechanism with auction-based credit pricing could possibly be implemented as follows \citep{liu2024tradable}: all travelers send their buying or selling requests containing the number of needed (or surplus) credits and the bidding (or offer) price to the market, which acts as a coordination platform (with additional functions) that sorts all requests, matches the buyers and sellers with transaction price determined by the matched orders (for example, the average value of the bid and offer prices \citep{worner2022bidding}), and subsequently sends the matching messages to related travelers. Upon receiving the matching message, sellers send credits to buyers and transaction records to the market. When the total credit supply is smaller than the demand, the regulator may be involved in transactions which it sells credits to buyers at a ceiling price that is higher than the highest offer price. The corresponding messaging process is similar to the one we describe in Figure \ref{sequence}.

\subsection{Demand model extensions}\label{sec:dem_ext}
The proposed demand model (pre-trip decision) is a combined departure time and route choice model. Assume that each traveler $n$ has a preferred arrival time $T_{n,i,d}^*$ for trip $i$ on day $d$. 
We assume the preferred arrival time is unchanged across days. The departure time choice set is individual-specific and defined as:
\begin{equation}\label{timewindow}
    \left.
    \begin{aligned}
         TW_{n,i,d}=\{\hat{t}_{n,i,d}-\eta\Delta_{tw},\hat{t}_{n,i,d}-(\eta-1)\Delta_{tw},\dots,\hat{t}_{n,i,d}+\eta\Delta_{tw}\},\\
    \end{aligned}
    \right.
\end{equation}

\noindent where $\eta$ is the size parameter, and $\hat{t}_{n,i,d}$ is traveler $n$'s preferred departure time interval on day $d$ for trip $i$. Thus, the departure time window (choice set) $TW_{n,i,d}$ consists of $2\eta+1$ time intervals of size $\Delta_{tw}$ centered around the preferred departure time, $\hat{t}_{n,i,d}$, which is the preferred arrival minus an estimated travel time. 
The size parameters of the time window ($\eta$) and time interval ($\Delta_{tw}$) are configurable. It is noteworthy that all activities and intermediate trips in the given trip chain from the \textit{Pre-day} module are performed in sequence, and hence a trip might be delayed by its previous trips. In other words, the end time of the previous activity, which is conducted after trip $i-1$ and before trip $i$, could be later than the edge of the time window of the current trip $i$, $\hat{t}_{n,i,d}-\eta\Delta_{tw}$. In this case, to maintain the number of departure time alternatives, we need to either shift the time window to the activity end time or shorten the activity duration or do both. This requires a more detailed model to account for the activity type (mandatory and discretionary) and the penalty of shifting the time window and shortening the activity. We defer this time window positioning problem to future research. In this study, we keep the activity duration unchanged and always shift the time window when needed.

The set of alternative paths between a given origin-destination node pair is generated via a combined method that uses the link elimination method, labeling method, k-shortest path method, and simulation method \citep{lu2015simmobility}. The choice set $\mathcal{K}$ of the combined departure time and route choice model is a product of the time window and path set. The combined departure time and route choice model uses the multinomial logit model. The utility $U_{k,n,i,d}$ and its systematic part $V_{k,n,i,d}$ of choosing alternative $k$ to traveler $n$ for a given trip $i$ on day $d$ are, respectively:
\begin{align}\label{5_utility}
&U_{k,n,i,d}=V_{k,n,i,d}+\epsilon_{k,n,i,d},\\
&V_{k,n,i,d}=\beta_{TT}\cdot TT_{k,n,i,d}+\delta_{n,i,d}\beta_{SDE,n}\cdot\big(T_{n,i,d}^*-t_k-TT_{k,n,i,d}\big)\\\nonumber
&+(1-\delta_{n,i,d})\beta_{SDL,n}\cdot\big(t_k+TT_{k,n,i,d}-T_{n,i,d}^*\big)+\beta_{cost,n}\cdot g(t_k)\cdot p_d\cdot Dist_k+\beta_{PS}\cdot P\_size_k\\\nonumber
&+\beta_{L}\cdot Dist_k+\beta_{sigNum}\cdot Sig\_num_k+\beta_{hwy_dist}\cdot Hwy\_dist_k+\beta_{minTT,k}\cdot Min\_TT_k\\\nonumber
&+\beta_{minDist}\cdot Min\_dist_k+\beta_{minSig}\cdot Min\_sig_k+\beta_{maxHwy}\cdot Max\_hwy_k,\\\nonumber
&k\in \mathcal{K},
\end{align}
where
\begin{description}
    \item $TT_{k,n,i,d}$: Travel time of choosing alternative $k$, which specifies the departure time $t_k$ and the route.
    \item $\delta_{n,i,d}$: Binary variable, $\delta_{n,i,d}=1$ when the arrival time is smaller the preferred arrival time, i.e., $t_k+TT_{k,n,i,d}<T_{n,i,d}^*$, otherwise $\delta_{n,i,d}=0$.
    \item $g(t_k)\cdot p_d\cdot Dist_k$: monetary value of the distance-based credit charge departing at $t_k$.
    \item $P\_size_k$: Path size the route from alternative $k$.
    \item $Sig\_num_k$: Number of signals in the route from alternative $k$.
    \item $Hwy\_dist_k$: Distance of the highway portion in the route from alternative $k$.
    \item $Min\_TT_k$: Does this route have the shortest travel time in the set.
    \item $Min\_dist_k$: Does this route have the shortest distance in the set.
    \item $Min\_sig_k$: Does this route have the minimum number of signals in the set.
    \item $Max\_hwy_k$: Does this route have the largest highway usage in the set.
    \item $\beta_{cost,n}$: Coefficient of toll cost, computed by $\frac{\beta_{TT}}{VOT_n}$, where the units are $\beta_{cost,n}$ (utility unit per minute) $\beta_{TT}$ (minute), and value of time (VOT, unit: \$/minute).
    \item $\beta_{sde,n}, \beta_{sdl,n}$: Coefficients of schedule delay early and late utility, computed by $\beta_{cost,n}\cdot SDE_n$ and $\beta_{cost,n}\cdot SDL_n$, respectively, where $SDE_n$ and $SDL_n$ (unit: \$/minute) are the monetary schedule delay early and late parameters.
    \item $\epsilon_{k,n,i,d}$: Identically and independently Gumbel distributed random component.
    \item $\beta_{TT},\beta_{L},\beta_{PS},\beta_{sigNum},\beta_{hwy_dist},\beta_{minTT,k},\beta_{minDist},\beta_{minSig},\beta_{maxHwy}$: Calibrated model parameters.
\end{description}
In this study, we do not consider nonlinear income effects, the allocation of credits is then a constant and does not affect the choice, thus it is not included in the utility function. 

In the mid-term \textit{With-day} module, a combined departure time and route choice is only made when the time approaches the time window of the corresponding trip. When applying the trading model in Section \ref{sec:MB}, the departure time of a future trip $i$, $t_{n,i,d}^{dep}$, is assumed to be the generated preferred departure time $\hat{t}_{n,i,d}$ unless the departure time is chosen. Thus the credit charge $g(t_{n,i,d}^{dep})$ and account balance $x_n^d(t_{n,i,d}^{dep})$ in equation \eqref{profit} are predicted values unless the time approaches the time window of the next trip, i.e., $t=\hat{t}_{n,i,d}-\eta\Delta_{tw}$. In addition, the selling model takes the trips on day $d$ and $d+1$ into consideration, wherein the latter are assumed to be duplicates of the former.

\subsection{Supply and day-to-day learning}
The supply simulator in SimMobility mid-term follows the design of DynaMIT supply model \citep{ben2002real,lu2015dynamit2}, which includes private cars and buses. The mesoscopic road traffic network in SimMobility mid-term is represented by a hierarchical structure composed of links, segments, lane groups, and lanes. Each link is composed of several segments which represent a section of homogeneous roadway, which is further divided into two traffic flow regions: a moving part and a queuing part. Vehicles in the moving part travel at some uniform positive speed determined by a predefined macroscopic speed-density function, while vehicles in the queuing part form a spatial queue whenever the arrival rate exceeds the segment capacity. The reader is referred to \citep{lu2015simmobility} for further details on the mesoscopic supply simulator of SimMobility Mid-term.

SimMobility mid-term has a two-level learning framework as shown in Figure \ref{midterm}. In the upper level (pre-day learning), the aggregate zone-to-zone traffic performance indicators, including travel cost, travel time, travel distance, and waiting time for different modes (e.g., car, public transit, on-demand services, etc.) feed back to the \textit{Pre-day} model to update agent's knowledge. In the lower level learning (within-day learning), the time-dependent link travel time can be iteratively updated considering a fixed \textit{Pre-day} demand and used in the \textit{Within-day} and \textit{Supply} in-simulation interaction when agents have to make the combined departure time and (link-level) route choices. Both pre-day and within-day loops stop until consistency is achieved \citep{basu2021framework}. In this study, we focus on the within-day day-to-day learning and keep the pre-day output (i.e., demand) fixed. An exponential smoothing filter is adopted to update the time-dependent link travel time, and the learning rate is the weight coefficient of the simulated (or realized) travel time. A larger learning rate leads to a more unstable system \citep{cantarella1995dynamic}.

\section{Bayesian optimization of TCS}
\label{sec:Opt}

Bayesian optimization (BO) is an efficient method for optimizing tolls in computationally intensive simulators like SimMobility, as introduced by \cite{liu2020bayesian} and implemented in SimMobility by \cite{Argyros2023} for congestion pricing. BO's main element is a model of the objective function and an acquisition function. 

\subsection{Bayesian Optimization formulation}

A Gaussian process (GP) regression is used to model the objective function as in equation (\ref{eq:objf}). GP regression is defined by a prior mean function $\mu(x)$ and a covariance function $\kappa(x,x')$, with $x,x'$ representing input variables related to the toll profile. Then, given a set of evaluated points, the GP is able to approximate the objective function, which in our case will be the social welfare (SW). 
The SW is calculated based on the observed individual travel utilities as in equation (\ref{eq:sw}) with reference to the no toll baseline ($U_{n,base}$), considering the initial free token allocation. The result is then divided by the cost coefficient to convert it into monetary units. 
\begin{equation} \label{eq:sw}
SW = \sum_{n=1}^{N}\frac{ ( U_{n} - \beta_{cost,n}\cdot g(t_k)\cdot p_d\cdot Dist_k) - U_{n,base}}{ |\beta_{cost,n}|}
\end{equation}


Thus, within BO, a GP is trained to approximate the SW:

\begin{equation} \label{eq:objf}
SW_{(x)} \sim GP(\mu(x), \kappa(x,x'))
\end{equation}

Moreover, several covariance functions to correlate points in the input domain can be used. A regular choice is the Mat\'ern kernel:
\begin{equation}  \label{eq:kernel}
k(x, x') =  \frac{2^{\nu-1}}{\Gamma(\nu)}\left( \frac{\sqrt{2\nu}}{l} |x - x'| \right)^\nu H_\nu\left(\frac{\sqrt{2\nu}}{l} |x - x'|\right)
\end{equation}
where $\Gamma(\cdot)$ is the Gamma function and $H_\nu$ is the modified Bessel function, with $\nu = 5/2$ used.

The acquisition function determines the next point to evaluate ($x_{next}$) based on the posterior mean function and the variance of the Gaussian Process (GP). A commonly used acquisition function is the Upper Confidence Bound (UCB) \citep{Srinivas2012} presented in equation (\ref{eq:aq1}). 
\begin{equation} \label{eq:aq1}
n_{(UCB)}(x,\rho) = -\mu(x) + \rho\sigma(x)
\end{equation}

The hyperparameter $\rho$ controls the balance between exploration and exploitation, with larger values of $\rho$ indicating greater exploration. In this context, $\rho$ equal to 2 was selected. Thus, the next point to evaluate is determined by maximizing the acquisition function as:
\begin{equation} \label{eq:aquaf}
x_{next} =  \operatorname*{argmax}_x n_{(UCB)}(x,\rho)
\end{equation}

\subsection{BO extension for SimMobility}
\label{BO_simmob}

\begin{figure}[htbp]
    \centering
    \includegraphics[width=0.85\textwidth]{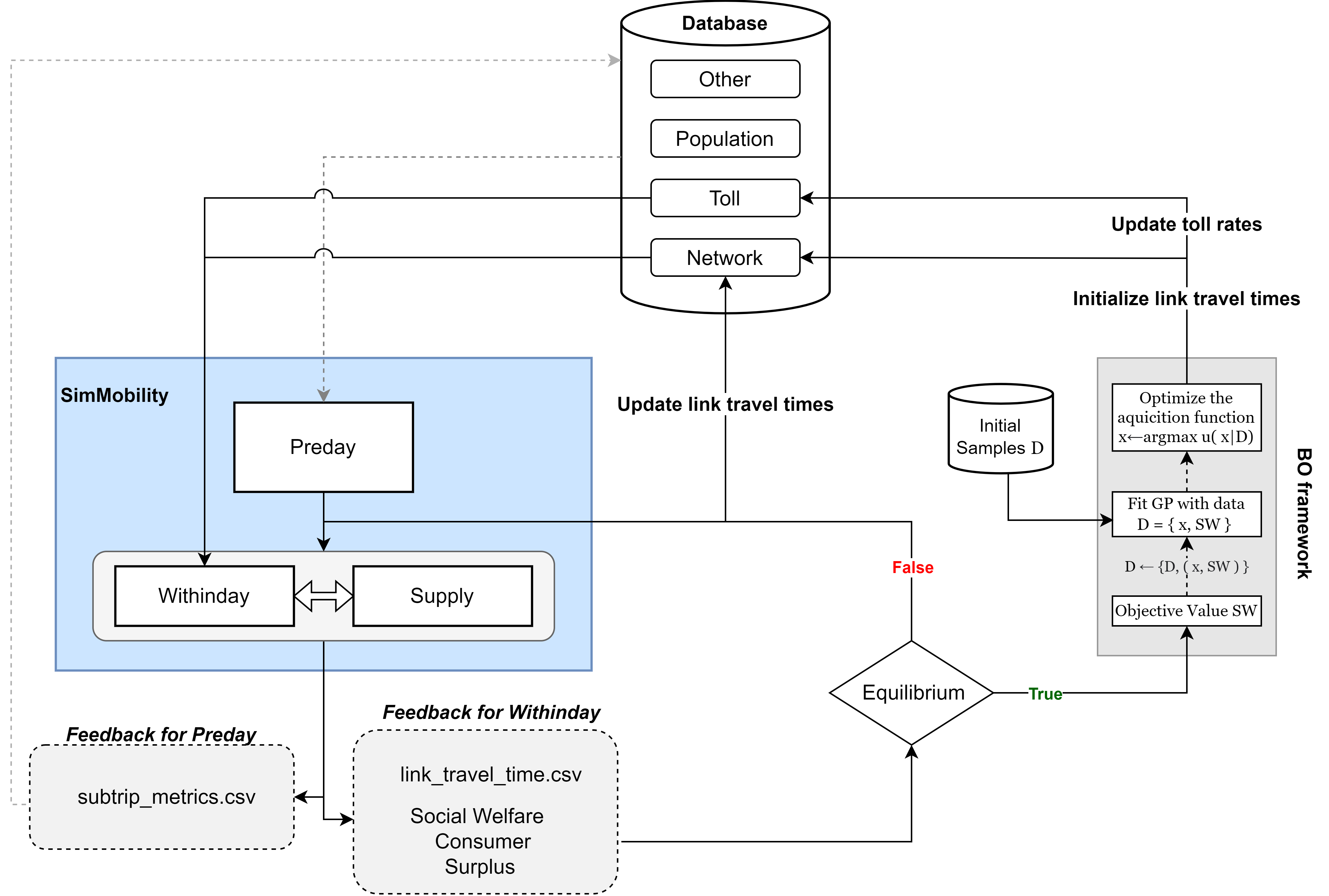}
    \caption{SimMobility and Bayesian optimization for road pricing}
    \label{bo_simmob_frame}
\end{figure}

The integration framework of Bayesian Optimization (BO) with SimMobility for tolling optimization is illustrated in Figure \ref{bo_simmob_frame} \citep{Argyros2023}. Here, we focus on within-day analysis, thus excluding the pre-day phase. In this framework, the knowledge about link travel times is updated and the social welfare is calculated for each within-day iteration. When consistency (equilibrium) is achieved, BO uses the average social welfare value from several within-day iterations as input for the GP and produces a new toll. A Gaussian curve is employed to generate the new distance-based, time-varying credit toll $g(t)$ (unit: credits/meter) in the form of a (non-negative) step toll whose rate varies every 5 minutes. Specifically, three variables are considered: the amplitude ($A_{g}$), which represents the highest toll price; the mean ($\mu_{g}$), which indicates the time at which this peak price occurs; and the standard deviation ($\sigma_{g}$) which defines the spread of the toll prices around the mean. Next, the framework initializes the link travel time knowledge in the database to the no-toll case and applies the new toll rates to the network. Lastly, set D represents an initial sample of tolls and their respective welfare values that are used to increase BO's efficiency and are updated for each new tested toll. 

\section{Experiment results}
\label{S:Results}
\subsection{Study area and experimental design}\label{sec:setting}
This case study is tested using the `Virtual City', which consists of a moderately sized network, generated so as to resemble land use patterns, travel behavior, and activity patterns observed in Singapore \citep{basu2018implementation,basu2021framework}, with calibrated parameters such as time-of-day, mode, destination choice, route choice, speed-density parameters, zone-to-zone travel time, etc. The road network consists of 95 nodes (intersections), 286 segments (road sections with homogeneous geometry), and 254 links (groups of one or more segments with similar properties). The population subset used contains 19,000 private car drivers. Among the daily activity schedule given by \textit{Pre-day}, 85\% of the trips are work-related, 10\% are education related and the rest are for other purposes (like shopping); on average, every person has 1.04 tours and 2.6 trips per day.

In this study, we perform 25 iterations (i.e., days) of the within-day learning for each toll evaluation to ensure that the system has reached stability. We investigate the within-day behaviors for an average weekday, including departure time and route choices and all TCS-related transactions. The pre-day daily activity schedule (by half-hour slots) is predetermined and fixed during the simulation \citep{basu2021framework}.

We generate the value of time using the log-normal distribution in \cite{chen2023market}, which is extracted from the Integrated Public Use Microdata Series (IPUMS) 2019 census data \citep{IMPUMS}. The average value of time is \$13/hour. The ratios of schedule delay early and late penalties to the value of time are assumed to follow triangular distributions with modes at 0.5 and 2, respectively, bounded by the widely used trip timing preferences relationship, i.e., the schedule delay early parameter is half of the value of time, which is in turn half that of the schedule delay late parameter \citep{vickrey1969congestion,small2012valuation}. The inference on these values under TCSs should be investigated, and we defer it to future research. Further,  Table \ref{tab:simpar} presents simulation parameters used for the TCS experiments. The presented values are only one feasible combination to showcase the functionality of the simulation framework, and their influence should be further explored in other TCS design assessments with our proposed framework in future work.

\begin{table}[htbp]
\caption{Simulation parameters for tradable mobility credits analysis
}
\label{tab:simpar}
\centering{%
\begin{tabular}{lr}
\hline
Variable description         & Value                     \\ \hline
Departure time window & \begin{tabular}[c]{@{}r@{}}60-minute time interval around \\ the preferred departure time\end{tabular} \\
Departure time interval      & 5 minutes                 \\
Day-to-day learning rate     & 0.2                      \\
Token allocation rate        & 1 credit every 20 minutes \\
Token lifetime               & 1420 minutes              \\
Initial credit price          & 0.1 \$                    \\
Initial credit allocation    & 72 credits                \\
Maximum credit cost per trip & 160 credits               \\
Transaction fees             & 0 \$                      \\ \hline
\end{tabular}%
}
\end{table}

\subsection{Base scenario}\label{sec:base}
We first examine the properties of the day-to-day process for our 'Virtual City', under no TCS. The free flow link travel time serves as the initial travel time information. The inconsistency between the predicted trip travel time (computed by the sum of historical link travel times covered by the chosen route) and the simulated trip travel time is used as a measure of convergence of the day-to-day evolution. The inconsistency is calculated as $\sum_n \sum_i^{I_n}|T_{n,i,d}^{\text{sim}}-T_{n,i,d}^{\text{pre}}|/\sum_n \sum_i^{I_n}T_{n,i,d}^{\text{sim}}$, where $T_{n,i,d}^{\text{sim}}$ is the simulated travel time of traveler $n$'s $i$th trip on day $d$, and $T_{n,i,d}^{\text{pre}}$ is the predicted travel time.

\begin{figure}[htbp]
    \centering
    \includegraphics[width=1\textwidth]{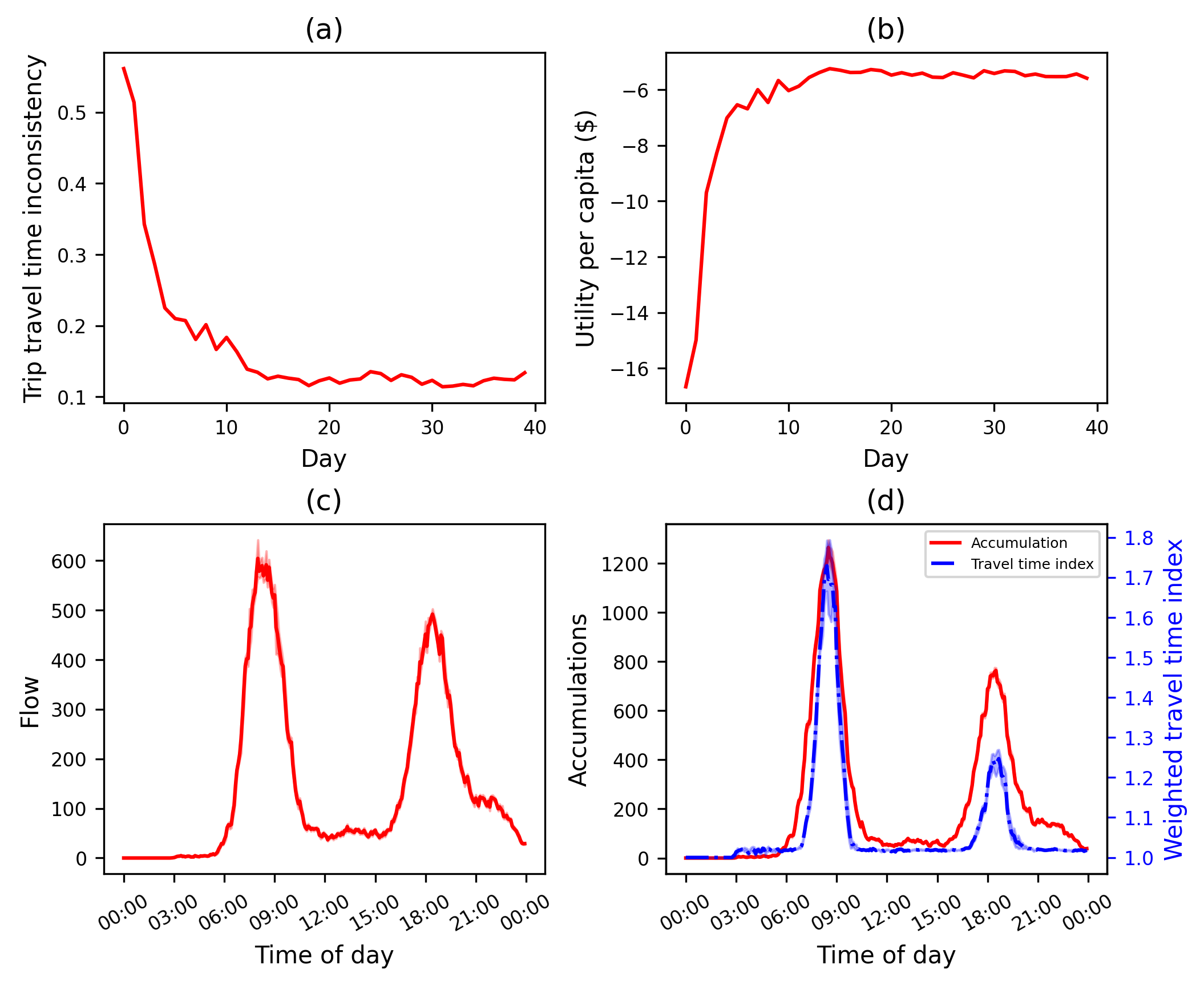}
    \caption{The evolution process with no toll}
    \label{base_fig}
\end{figure}

Figure \ref{base_fig}(a) presents the convergence of the trip travel time. It is found that the inconsistency becomes stable around 0.15 after day 15, implying that the day-to-day evolution process achieves stationarity and an acceptable degree of consistency. Figure \ref{base_fig}(b) shows the evolution process of the experienced utility per capita. Again, convergence is observed with small variability across iterations. We perform a commonly used stationary test, the Augmented Dickey-Fuller (ADF) test \citep{cheung1995lag}, on the utility data to examine the stationarity. If the p-value of the test is smaller than 0.05, the null hypothesis is rejected, i.e., the tested data does not have a unit root and it is called a stationary series. In this experiment, we have a p-value $\ll$ 0.05 hence stationarity is achieved. Figure \ref{base_fig}(c) illustrates the (car-based) departure rates for all activity purposes by 5-minute intervals of an average day, clearly displaying the morning and evening peak periods, wherein the curve shows the average values across the last 10 days, and the shadow stands for the 95\% confidence interval We also plot the accumulation and weighted travel time index to quantify the extent of congestion on the network. Accumulation is simply the number of vehicles on the network at a certain time interval. The travel time index (TTI) is the ratio of realized trip travel time to free-flow travel time.
The TTI is weighted by trip distance for every 5-minute interval and computed as: 
\begin{equation}
   TTI_{\Tilde{t}}=\frac{\sum_{i}\frac{Dist_i\cdot TT_i}{TT_i^{ff}}}{\sum_{i} Dist_i}, \forall i~\text{that} ~\Tilde{t}-1<t_i<\Tilde{t}, \nonumber
\end{equation}
where 
\begin{description}
    \item $t_i$: departure time of trip $i$,
    \item $\Tilde{t}$: time interval,
    \item $Dist_i$: distance of trip $i$,
    \item $TT_i$: experienced travel time of trip $i$,
    \item $TT_i^{ff}$: free-flow travel time of trip $i$.
\end{description}
In Figure \ref{base_fig}(d), the red curve represents the average accumulation across the last 10 days at 5-minute resolution and the shadow area the 95\% confidence interval are presented. Similarly, the blue curve represents the weighted TTI, with peaks of 1.8 in the morning and 1.3 in the evening. In \cite{Argyros2023}, where an identical basecase used, a toll that improves social welfare was only found to exist during the morning peak. Thus, here our focus is on utilising BO to design a distance-based toll that targets the morning peak.  

\subsection{TCS scenario}\label{sec:trinity}
\begin{figure}[htbp]
    \centering
    \includegraphics[width=1\textwidth]{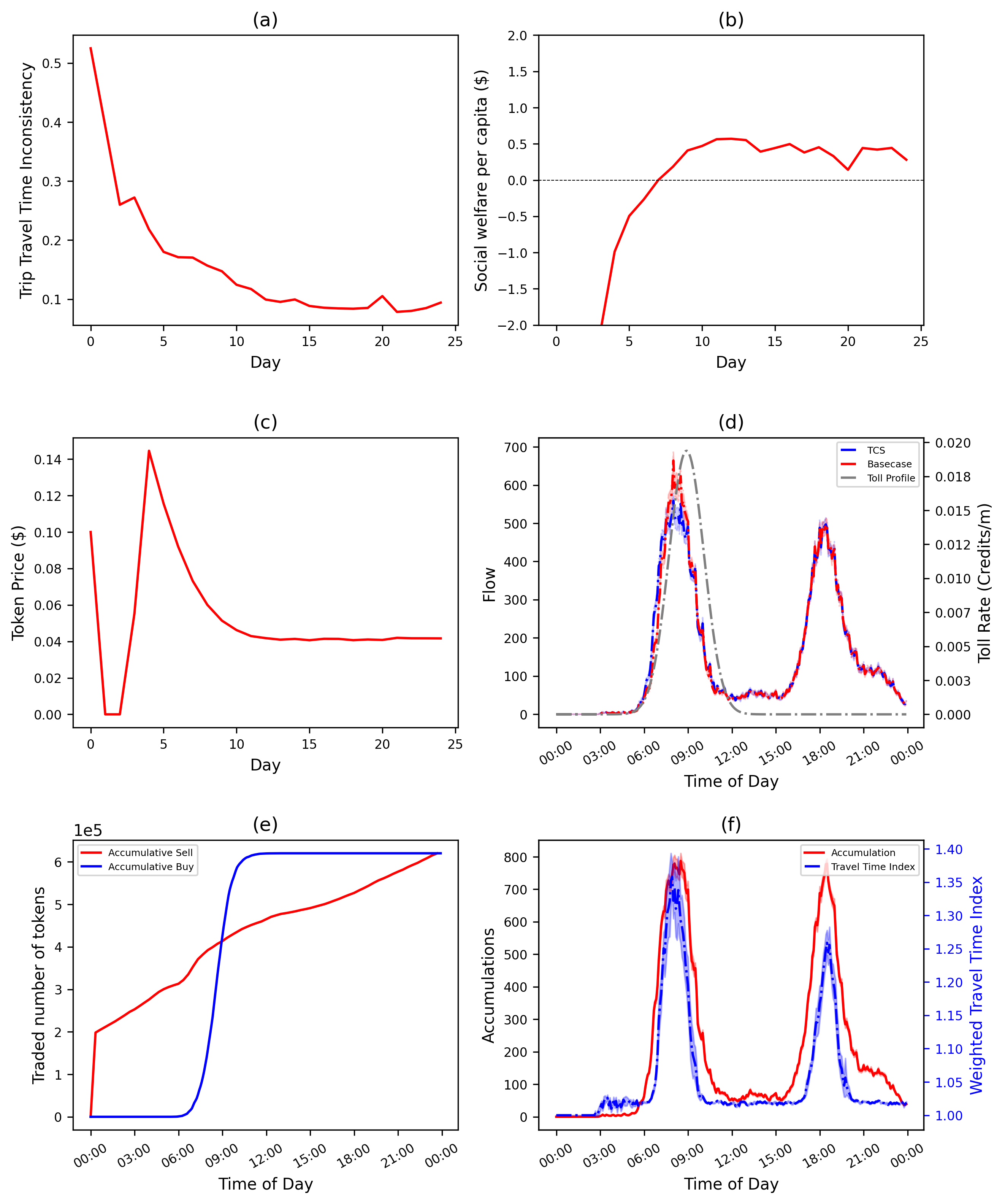}
    \caption{The evolution process with a distance-based credit toll}
    \label{trinity_fig}
\end{figure}

In this section, we present the simulation results under a distance-based credit step tariff. To design the tariff, we used Bayesian optimization which ran for 30 toll iterations. We then chose the one that yielded the highest social welfare. As presented in Section \ref{BO_simmob} a Gaussian curve is used to generate the toll. Here the optimal toll has a peak ($\mu_{g}$) at 8:59 in the morning, with an amplitude of 0.0194 credits per kilometre and 72 minutes standard deviation. Utilizing this profile we generate the tariff for each five-minute interval. The evolution process is shown in Figure \ref{trinity_fig}.

\begin{figure}[htbp]
    \centering
    \includegraphics[width=0.85\textwidth]{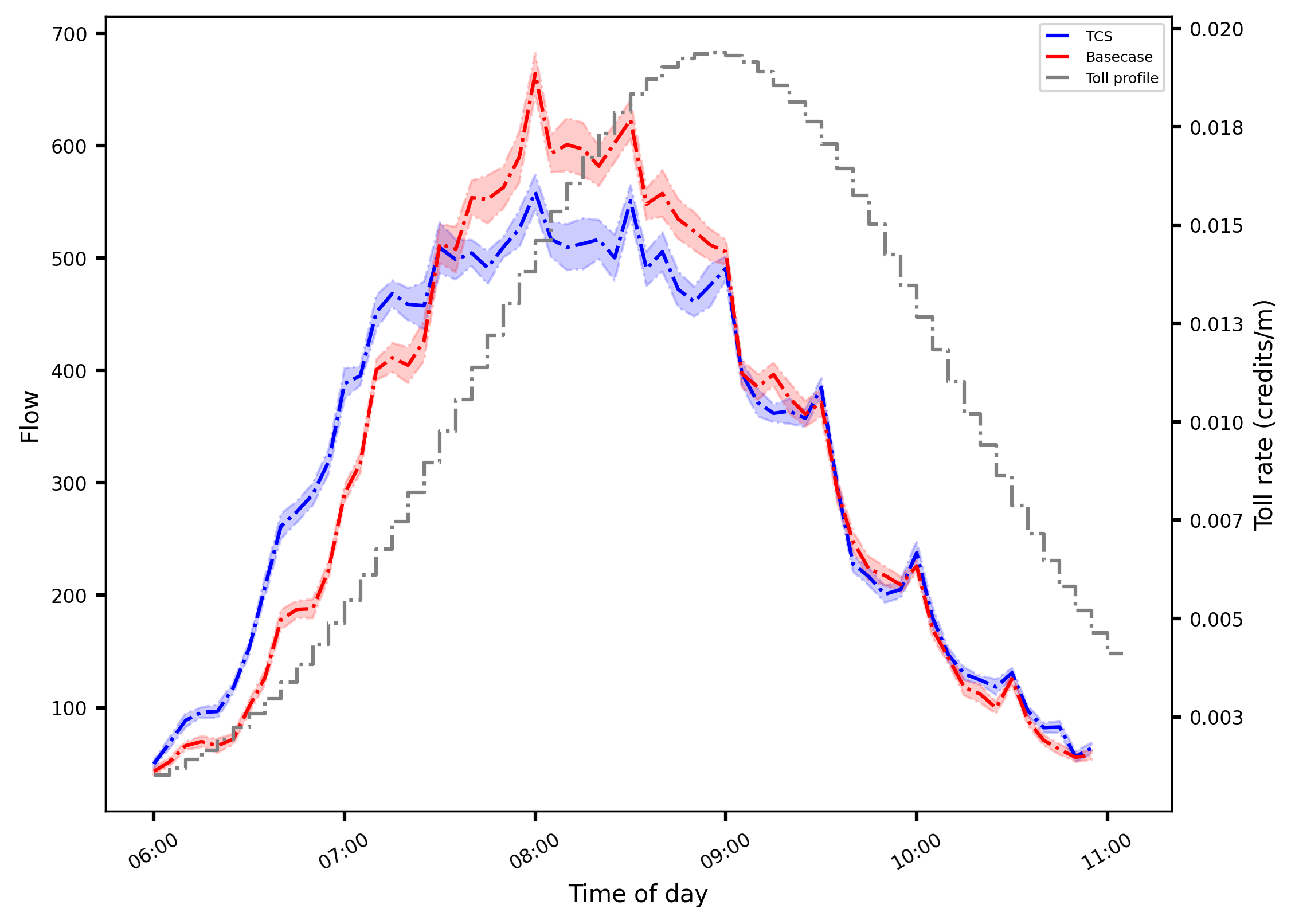}
    \caption{The flows of the morning peak}
    \label{morning}
\end{figure}

The inconsistency of trip travel time reduces and becomes stable around 0.1 after day 15, as presented in Figure \ref{trinity_fig}(a). We plot the social welfare gains per capita in Figure \ref{trinity_fig}(b), which is computed as the average difference of the simulated individual specific utilities between the TCS case and base case. The average social welfare gains per capita (driver) of the last 10 days is \$0.38, with a 95\% confidence interval of [\$0.3, \$0.46]. Figure \ref{trinity_fig}(d) shows the flows (red curve for base scenario and blue curve for TCS scenario) and tariff profile (colored in grey). We plot the flows of the morning peak for better illustration in Figure \ref{morning}. Clearly, the flow curve is flattened as the demand shifts to earlier intervals to avoid the larger credit charges. Consequently, as shown in Figure \ref{trinity_fig}(f), the peak accumulation reduces from around 1200 vehicles to 800 vehicles, and congestion is mitigated as reflected by the reduction of the weighted TTI from 1.8 to 1.35, leading to gains in social welfare. Figure \ref{trinity_fig}(c) displays the evolution of the credit price, which stabilizes at \$0.041. The result is consistent with the evolution of the credit transactions on an average day plotted in Figure \ref{trinity_fig}(e), where the numbers of sold and bought credits reach the same level. The jump in the curve of accumulative buying transactions stand for the morning peak of departures in the flow. 

\begin{figure}[htbp]
    \centering
    \includegraphics[width=0.85\textwidth]{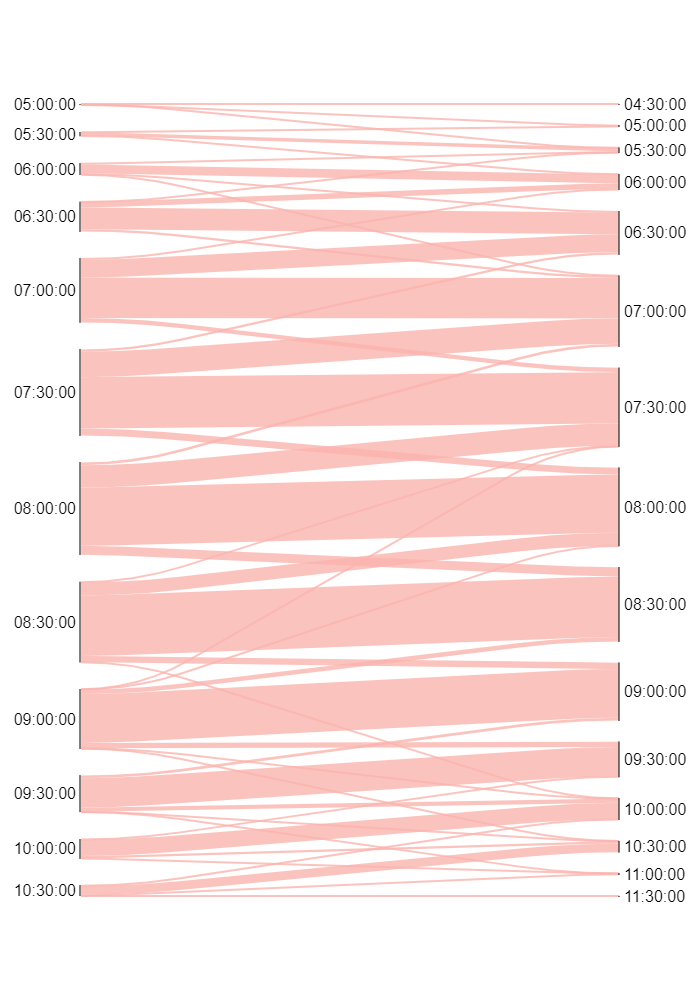}
    \caption{Changes in departure time, from base (left) to TCS (right) scenarios}
    \label{base_TCS}
\end{figure}

In addition to Figure \ref{morning}, we further plot the individual shifts in departure time in Figure \ref{base_TCS} from the base scenario to the TCS scenario. Times on both sides of the figure represent the trip start time of a 30-minute interval (e.g., 06:30:00 represents the time interval [6:30,7:00]). The widths of an arc or time-interval are proportional to their quantity with respect to the total demand. Before the peak of the credit tariff, i.e., before 9:00, it is observed that arcs connecting to an early time interval are thicker compared to those connecting to a later interval, indicating that most behavior shifts are earlier departure choices, avoiding higher credit charges. Similarly, after the tariff peak (after 9:30), more travelers postpone their trips.

\begin{figure}[htbp]
    \centering
    \includegraphics[width=1\textwidth]{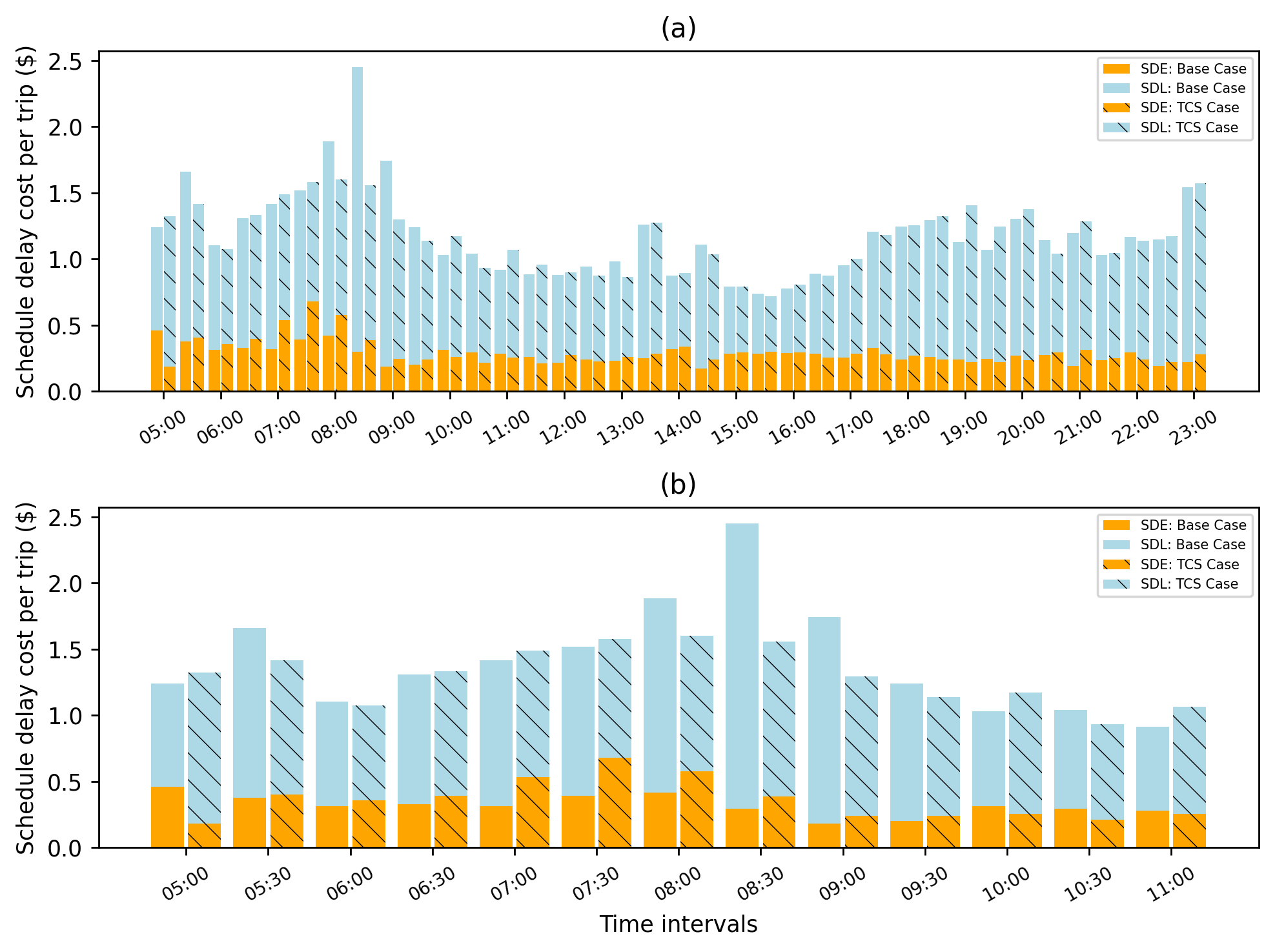}
    \caption{Schedule delay cost in base and TCS scenarios. Full day (top) and Morning period only (bottom)}
    \label{sd}
\end{figure}

We also evaluate the changes in schedule delay costs for both the base and TCS scenarios in Figure \ref{sd}(a) for the full day and Figure \ref{sd}(b) for the morning peak. The $x$-axis represents the 30-minute interval of the preferred arrival time, and the $y$-axis is the schedule delay cost per trip in the corresponding time interval. Note that the average trip travel time is 7.7 minutes and 95\% of trip travel time is smaller than 17 minutes, hence most trips start and end within the same 30-minute interval. Before the tariff peak (8:00), the sums of the schedule delay early and schedule delay late costs in TCS scenario are close to those in the base case, while the proportion of schedule delay early clearly increases, due to earlier departure times. During the peak, schedule delay late cost is significantly reduced thanks to the savings in travel time. After the peak (9:30), though we observe a few trip postponements in both Figure \ref{morning} and Figure \ref{base_TCS}, the schedule delay late costs do not increase significantly as travel time is reduced.

\begin{figure}[htbp]
    \centering
    \includegraphics[width=1\textwidth]{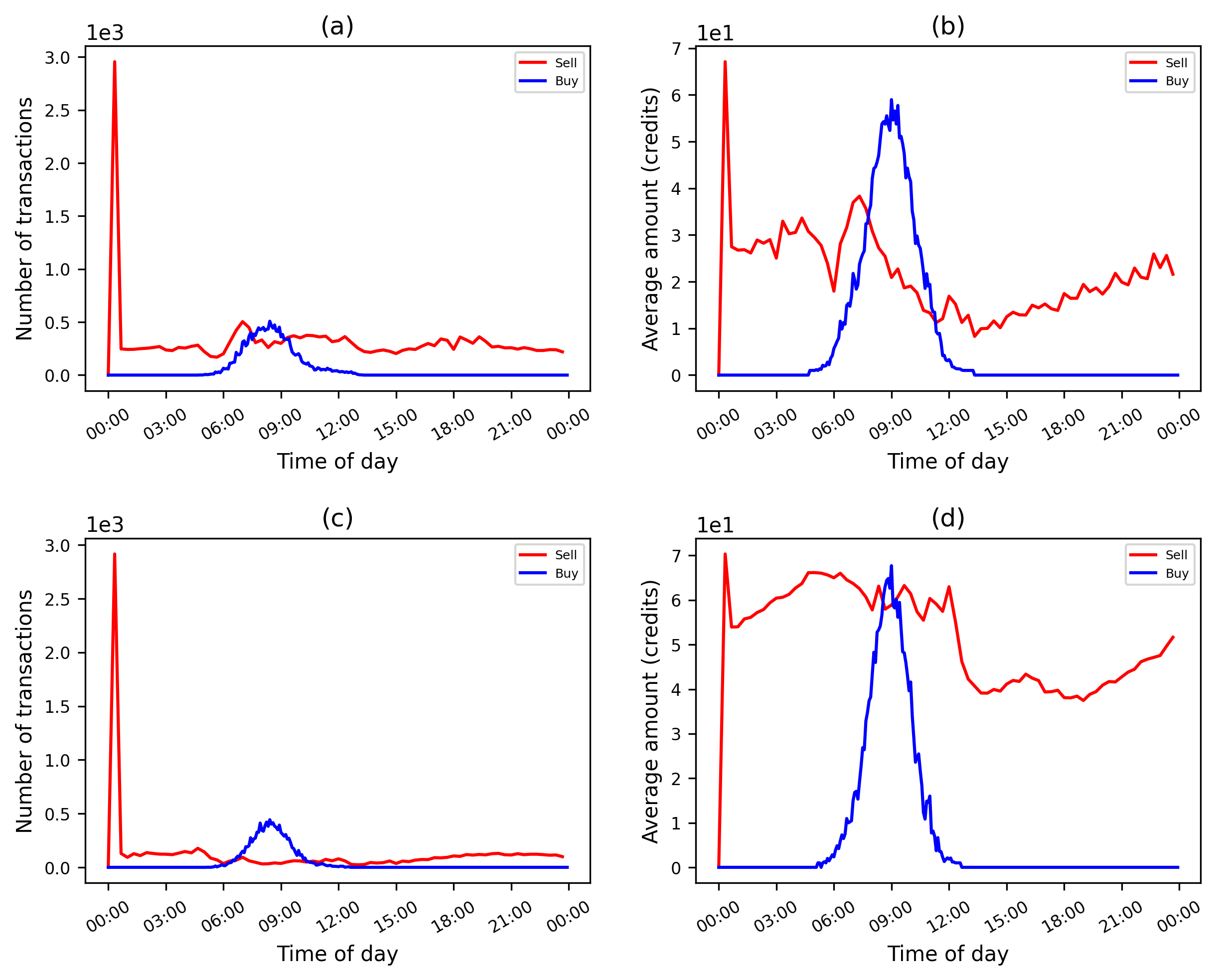}
    \caption{The transaction numbers (left) and average trading amount (right) with (bottom) and without (top) a profit threshold}
    \label{trinity_trans}
\end{figure}

Another critical characteristic of a TCS system that needs to be investigated is trading behavior in the market. As shown in Figure \ref{trinity_trans}(a), buying transactions only happen in the peak hour since travelers can only buy credits at the time of departure if they are short of credits. In contrast, the selling transactions happen across the day. For travelers who sell at the beginning of the day, most of them sell at full wallets as shown in Figure \ref{trinity_trans}(b), where the average trading amount is close to the full wallet (72 credits). In addition, relatively large amount of selling is observed during the peak hours. This is because of the selling model which requires a predicted toll that may be inaccurate because the chosen path and departure time is not known before the choice is made. For most of the trips, the realized credit charge is smaller than the predicted one, thus the selling profit becomes positive after the departure time is chosen, triggering a selling decision according to the selling model in Section \ref{sec:MB}. The other instances of selling are scattered across the day with an increase towards the end of the day. Fluctuations occur due the tariff profile, which has multiple (5-minute) steps and that the tariff step is not a multiple of the credit allocation per time interval. In the work of \cite{chen2023market}, travelers have just a single trip, a high resolution (every minute) of credit allocation, and a toll with five steps. There, the selling pattern observed here (high number of transactions  at the beginning of the day, the early morning, and during peak hours) is also present, but with fewer fluctuations since the factors mentioned above are absent.

\begin{figure}[htbp]
    \centering
    \includegraphics[width=1\textwidth]{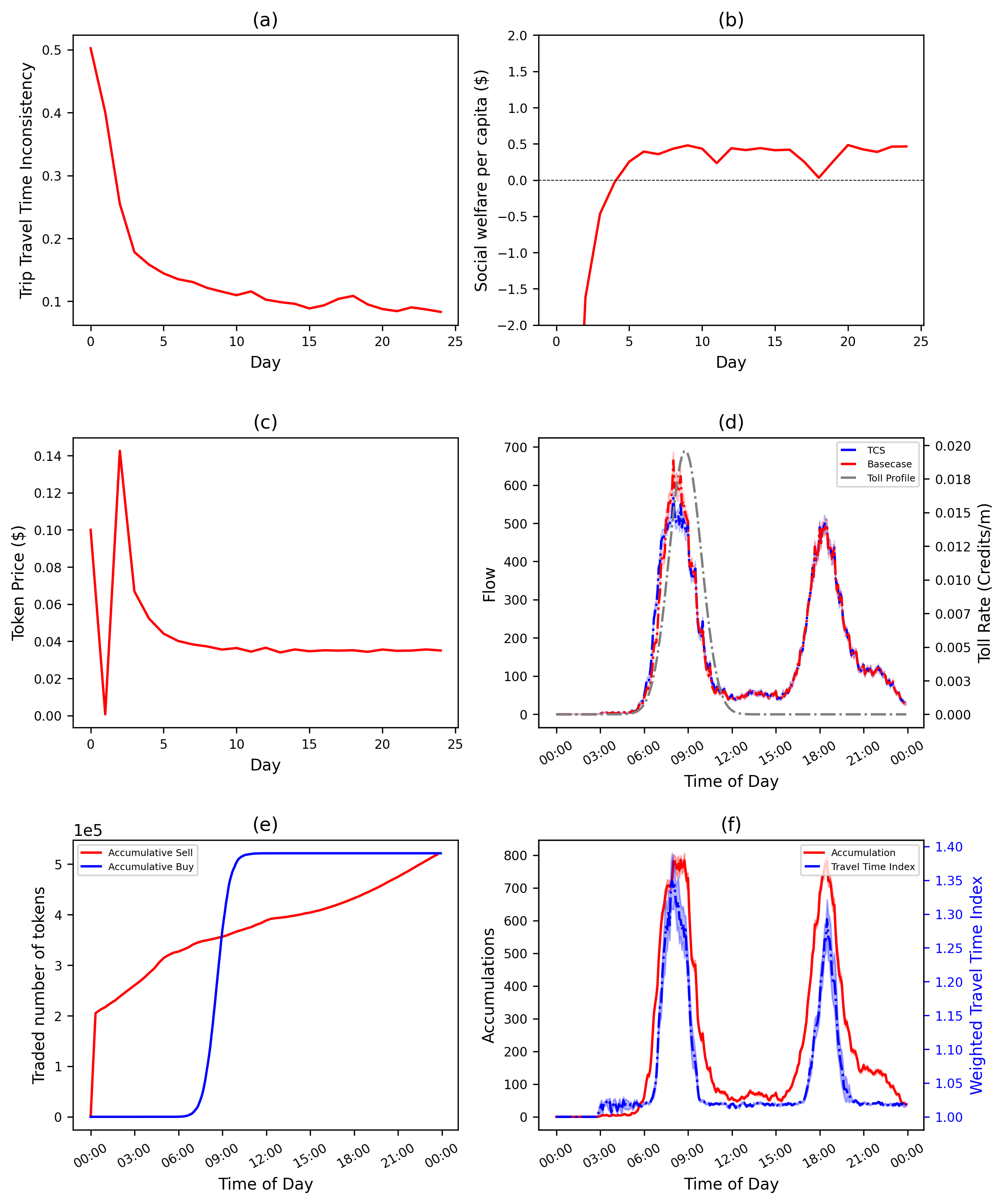}
    \caption{The evolution process of TCS with a profit threshold}
    \label{threshold}
\end{figure}

Moreover, it is found in Figure \ref{trinity_trans}(b) that there are many selling transactions with small amounts of credits (smaller than 20 credits), as well as the phenomenon of selling followed by subsequent buying (termed buyback). This is undesirable because if the credit system is used by certain individuals to make a profit via speculation, it could hamper the acceptability of such a system. To prevent such undesirable market behaviors, \cite{brands2020tradable} propose to use a small transaction fee to suppress the frequent selling. 
Alternatively, we propose to set up a threshold on the selling profit. In particular, travelers are only allowed to sell credits if the current selling profit is larger than the predetermined threshold. A threshold of \$1 is adopted in this study and the BO is used again to design the optimal toll profile. A similar toll was detected ($\mu_{g}:$ 8:54, $A_{g}:$ 0.0196 and $\sigma_{g}:$ 66 ) and the evolution process under the new tariff is shown in Figure \ref{threshold}. Besides the convergence, we observe similar social welfare gains, departure flow, accumulation and weighted TTI. The stable credit price is \$0.035, which is lower than that of the case without the profit threshold, caused by the changes in toll shape and selling behaviors and the resulted departure time choices.
In Figure \ref{threshold}(e), it is worth noting that the total number of traded credits is significantly reduced compared to that in Figure \ref{trinity_fig}(e).

\begin{figure}[htbp]
    \centering
    \includegraphics[width=1\textwidth]{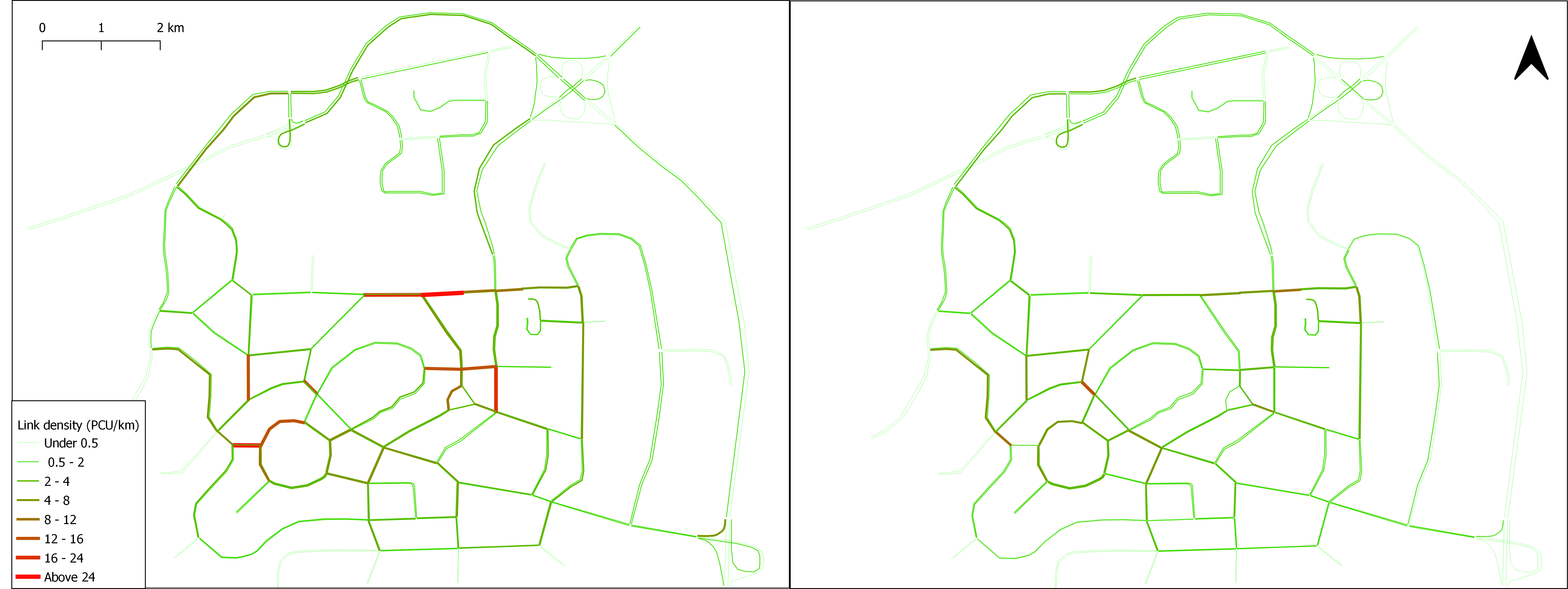}
    \caption{Average network density improvement from the basecase (left) to the TCS case with profit threshold (right) during the peak hour (8:00 - 9:00) for an average day at stability.}
    \label{network}
\end{figure}

Leveraging the detailed network of our simulator, we illustrate the improvement achieved from the introduction of a TCS scheme in the network’s density level in Figure \ref{network}, with the toll clearly alleviating network congestion bottlenecks. Lastly, we summarize the relevant factors such as numbers of transactions, traded credits, and travelers with buyback behavior, social welfare gain, and credit price in Table \ref{trinity_compare}. As desired and expected, after imposing a small profit threshold, the number of transactions is greatly reduced, and fewer travelers have the buyback behavior. Also, the average trading amount of credits is found to be larger, as shown in Figure \ref{trinity_trans}(d) and (b). Social welfare is similar, but slightly decreased, compared to the no profit threshold case. One may expect that the resulting lower credit price could cause a fewer shift of departures in periods with relatively lower tariffs, however here the effect of the threshold in the social welfare gains seems to be minimal. Nevertheless, imposing a profit threshold is an effective way to control undesirable trading behaviors. More comprehensive experiments are needed to test the price elasticity and compare the performance of transaction fees and the profit threshold. Lastly, for verification purposes, we conducted three replications of the threshold case using different random seeds in the simulator. The resulting standard deviation was small, indicating that the outcomes were consistent across the runs, with similar values observed in each replication.

\begin{table}[!ht]
  \centering
  \caption{Statistics of the TCS with and without profit threshold}\label{trinity_compare}
  \vskip 0.2cm
  \begin{tabular}{lll}
\hline\noalign{\smallskip}
Factors & Without & With \\
\noalign{\smallskip}\hline\noalign{\smallskip}
Profit threshold & 0 & \$1 \\
Sell No. Trans. &22780 &8992 \\
Buy No. Trans. &16852 &12173  \\
Traded credits &1.24$\times 10^6$ & 1.04$\times 10^6$\\
Travelers with buyback &4881 &1651 \\
Welfare gain &\$0.38($\pm 0.08$) &\$0.36($\pm 0.1$)  \\
Credit price &\$0.041 &\$0.035 \\
Tariff profile &(8:59, 72, 0.0194) &(8:54, 66, 0.0196) \\
\hline
\end{tabular}
\end{table}

\section{Conclusions}
\label{5_S:Con}
This paper proposes the first flexible simulation framework with a modular and extensible implementation in a state-of-the-art agent- and activity-based simulator for the detailed modeling and assessment of TCS. Demand is modeled through an activity-based model and a within-day departure and route choice model sensitive to individual credit charges and heterogeneous preferences. The transportation supply is represented by a multi-modal mesoscopic network model and is extended with TCS capabilities for handling all credit transactions within the simulation. This proposed framework allows for the simulation of a variety of TCS design schemes and a selected recent TCS design was here tested and its properties validated for showcasing in a prototypical urban setting.

The results show convergence of the day-to-day process in terms of trip travel time, credit price, flow pattern, and travel utility and validate the general properties of TCS previously demonstrated in theoretical models. As expected, and thanks to the suitability of BO for complex simulation-based optimization, an optimised credit tariff triggers the demand shift in peak hours to alleviate the overall congestion and improve social welfare compared to 11the no-toll case. Moreover, the results suggest that a small profit threshold is able to effectively control undesirable trading behaviors such as frequent selling and buyback.

It should be pointed out that the current TCS results, including the welfare gain estimates, do not account for behavior changes from the \textit{Pre-day} activity-based model, thus not accounting for possible shifts in activity types, number, duration, time-of-day, destination and mode choices, which constitutes an important avenue of future assessment using our proposed framework. Also, with our proposed simulation framework, modelling complex market designs, individual market behaviors and networks are now possible, thus significantly contributing to the assessment of TCS properties and impacts of alternative demand management mechanisms \citep{liu2024tradable} and heterogeneous market behaviors \citep{dogterom2017tradable}. Such alternative designs and possible behaviors should be tested. Finally, the performance of TCS under demand and supply variability and uncertainty is also an avenue for exploration with our proposed framework.






\bibliographystyle{elsarticle-harv}
\bibliography{ref.bib}







\end{document}